\documentclass[aps,pra,twocolumn]{revtex4-2}

\pdfoutput=1
\usepackage{amsmath,amssymb}
\usepackage{times,txfonts}
\usepackage{graphicx}
\usepackage{hyperref}

\graphicspath{{../../figures/}}

\newcommand{\vphast}{{\vphantom *}}

\newcommand{\bk}{\mathbf{k}}
\newcommand{\br}{\mathbf{r}}

\renewcommand{\Im}{\mathop{\mathrm{Im}}}
\renewcommand{\Re}{\mathop{\mathrm{Re}}}

\begin{document}

\date{February 16, 2026}

\title{Phase domain walls in coherently driven Bose-Einstein
  condensates}

\author{S.~S.~Gavrilov}

\affiliation{Institute of Solid State Physics RAS, 142432
  Chernogolovka, Russia}

\begin{abstract}
  We consider coherent states of weakly interacting bosons under the
  conditions of external resonant excitation, with a focus on a
  two-dimensional polariton fluid driven by a plane electromagnetic
  wave near the ground state.  The coherent driving breaks the U(1)
  symmetry explicitly, which prevents the occurrence of quantum
  vortices in a uniform scalar condensate.  Surprisingly, a spinor
  (two-component) system of the same kind admits topological
  excitations, such as domain walls of relative phase or confined
  half-vortex molecules, typical of a freely evolving spinor Bose
  system.  Opposite-phase domains arise from the spontaneous breakdown
  of the spin symmetry $(\mathbb{Z}_2)$.  Domain walls form with time
  even when the initial state of the system is uniform or completely
  disordered; they fall into different topological types distinguished
  by the total phase variation in the transverse direction.  One type
  of domain walls is similar to ``magnetic'' solitons in Bose-Einstein
  condensates and exhibits nonzero spin polarization whose sign
  depends on the direction of motion.  Domain walls of the second
  type, by contrast, behave like monopoles with broken spatiotemporal
  symmetry and tend to move in certain preferred directions.  The
  interaction of vortices and domain walls results in a long-range
  ordering of the system.
\end{abstract}

\maketitle

\section{Introduction}
\label{sec:introduction}

A coupled pair of spatially overlapping Bose-Einstein condensates
(BECs) has complex topological excitations that cannot be reduced to
the case of one scalar BEC.  Typical examples of such systems include
atomic BECs with coexistent hyperfine species~\cite{Leggett2001} and
two-band superconductors~\cite{Tanaka2001}.  In one dimension, they
admit the sine-Gordon kinks characterized by the $2 \pi$ jump of the
relative phase of order parameters~\cite{Son2002,Tanaka2001}; in two
dimensions, the kink turns into a domain wall which gives birth to
more complicated objects such as bound states of half-quantized
vortices (HQVs)~\cite{Eto2011,Tylutki2016}.  The mechanism of the HQV
pairing on a domain wall was found to be formally similar to quark
confinement~\cite{Eto2018,Eto2020}.  The underlying mean-field theory
attracts much interest owing to very rich physics it contains, in
spite of conceptual simplicity, and a broad scope which includes
BECs~\cite{Recati2022}, superfluid liquids~\cite{Makinen2019},
superconductors~\cite{Tanaka2015}, and even
cosmology~\cite{Kibble1982,Rubakov1983}.

In this work, we investigate a dissipative BEC coupled to an external
classical field whose frequency is close to the BEC ground-state
level.  Such a system is found to admit stable domain walls and HQV
molecules which arise spontaneously even when the initial conditions
are uniform or, conversely, disordered.  A uniform BEC undergoes a
spontaneous $\mathbb{Z}_2$ symmetry breaking that proceeds
\emph{nonglobally,} by way of the Kibble-Zurek
mechanism~\cite{Kibble1980,Zurek1985,Kang2019}, as soon as the driving
field reaches a critical amplitude; a pair of the true (asymptotically
stable) vacuum states of the system have opposite phases.  On the
other hand, in a case of disorder the system exhibits
self-organization.  Once dynamic equilibrium has established, a number
of HQV molecules and domain walls act as elements of complex,
inhomogeneous yet internally ordered states whose spatial extent is
far beyond any conventional characteristic scale of a~BEC.

The minimum model containing the phenomena under consideration is
represented by the mean-field equations
\begin{equation}
  \label{eq:mean-field}
  i \hbar \frac{\partial \psi_\pm}{\partial t} =
  -i \gamma \psi_\pm + \frac{\delta H}{\delta \psi_\pm^*}
\end{equation}
for a pair of complex-valued amplitudes $\psi_\pm = \psi_\pm(\br, t)$,
where $\gamma$ is the decay rate and
\begin{multline}
  \label{eq:energy}
  H \left[ \psi_+, \psi_- \right] =
  \int \, \Bigl[
  - D \left(
    |\psi_+|^2 + |\psi_-|^2
  \right)
  + \frac{\hbar^2}{2 m} \left(
    |\nabla \psi_+|^2 + |\nabla \psi_-|^2
  \right)
  \\ {} + \frac{V}{2} \left(
    |\psi_+|^4 +|\psi_-|^4
  \right)
  + \frac{\Omega}{2} \left(
    \psi_+^* \psi_-^\vphast +
    \psi_-^* \psi_+^\vphast
  \right)
  \\ {} + f \left(
    \psi_+^\vphast +
    \psi_-^\vphast +
    \psi_+^* +
    \psi_-^*
  \right)
  \Bigr] \, d^2 \br
\end{multline}
is the energy of a bosonic fluid characterized by mass $m,$ nonlinear
interaction strength $V,$ and spin coupling rate $\Omega$.  The
external field has amplitude $f$ that is assumed to be real-valued and
constant; thus, $D$ equals the detuning between the excitation energy
level (taken as zero) and BEC ground state $(-D)$.  In what follows,
we suppose that $V > 0$ and $D > \gamma$ and that feasible values of
$V|\psi_\pm|^2$ can be as great as~$D$.

This theory applies to a cavity-polariton system driven by a resonant
light wave~\cite{Kavokin.book.2017,Gavrilov2020.usp.en}.  Cavity
polaritons are bosonic particles, mixed states of excitons and photons
formed in a planar quantum well inside a microcavity.  They reach
macroscopic coherence either through Bose-Einstein condensation of
excitons~\cite{Kasprzak2006} or, thanks to the photonic part, under
coherent optical driving~\cite{Baas2006}.  Two pseudospin states of
excitons match the right- and left-handed circular polarizations of
light, which gives two polariton components.  The polariton-polariton
interaction $({\propto} \, V)$ comes from the exchange Coulomb
interaction of electrons or holes~\cite{Ciuti1998,Vladimirova2010}; an
inter-component scattering of this kind is usually
inhibited~\cite{Sekretenko2013.10ps} and disregarded
in~\eqref{eq:energy}.  At the same time, the two components can be
linearly coupled $({\propto} \, \Omega)$ owing to the structural
anisotropy of a microcavity~\cite{Gavrilov2014.prb.j}.  The decay rate
$\gamma$ depends on the cavity quality factor.

Apparently, the coherent driving alters the nature of a BEC.  The
model has the explicitly broken phase symmetry, i.\,e., it is not
invariant under the transformation
$\psi_\pm \mapsto \psi_\pm e^{i \chi}$ ($\chi$ being a real number)
unless $f = 0$.  It might seem unclear if this model ever admits
vortices or dark solitons.  In fact, they are absent in a scalar
(spinless) system driven by a continuous wave with zero in-plane
momentum, because the local phase of $\psi$ is fixed by the driving
field~\cite{Pigeon2011}.  The spinor model was found to admit both
solitons and vortices, but their relation to the ones typical of the
freely evolving BECs remained obscure~\cite{Gavrilov2020}.  Here, we
show that a nontrivial crossover exists between the driven and free
systems where they get close to each other in the sense of their
vacuum states as well as the elementary and topological excitations.

Our general predictions are the most easily understood in the limit
$\gamma \ll D \sim \Omega$.  Owing to the symmetry breaking, in a wide
range of $f$ the system has only two equally feasible vacuum states
$\psi_{\pm(\mp)}^\vphast = \psi_{\mp(\pm)}^* = \sqrt{n} e^{\pm i
  \alpha}$, where $n$ is constant and $\alpha$ grows from $\pi / 2$ to
$\pi$ with increasing $f$.  These solutions differ in the sign of the
relative phase $\arg (\psi_+^\vphast \psi_-^*) = \pm 2 \alpha$ and can
be interfaced by a domain wall.  The individual phases of $\psi_\pm$
inside both domains are determined modulo $2 \pi$, so the domain walls
fall into distinct topological types.  For instance, when $\psi_+$ or
$\psi_-$ gets switched from $\psi \sim e^{i \alpha}$ to
$e^{-i \alpha}$, the respective phase $\phi = \alpha$ can either
monotonically decrease down to $-\alpha$ or increase up to
$2 \pi - \alpha$; thus, it varies by $-2 \alpha$ or
$2 \pi - 2 \alpha$, whereas the other spin component has to change its
phase by $2 \alpha$ or $2 \alpha - 2 \pi$.  As a result, the amount
$q$ of the total continuous variation of $(\phi_+ + \phi_-) / 2 \pi$
across a domain wall takes discrete values $-1$, 0,~1.  Dynamical
instability of domain walls with $q = 0$ involves spontaneous
formation of confined HQV pairs, in analogy to atomic
BECs~\cite{Ihara2019,Gallemi2019,Gallemi2016}.  Second, a uniform
motion of the $q = 0$ domain wall is counterbalanced by a nonzero spin
polarization $(\propto n_+ - n_-)$ whose sign is fixed by the
direction of motion.  A similar phenomenon in atomic BECs, referred to
as a \emph{magnetic} domain wall or soliton, was also considered
recently~\cite{Qu2017,Yu2021,Yu2022}.

By contrast, the $q = \pm 1$ domain walls are quite unusual objects
with fundamentally broken spatial and spin symmetries.  Two phase
gradients $\nabla \phi_\pm$ now have the same direction that is
determined by $q$ irrespective of the actual solitonic velocity.  The
spin polarizations and energetically favored directions of motion are
opposite for the $q = \pm 1$ domain walls which, thus, cannot
transform into each other.  At the same time, they can be linked
together by a HQV pair and make up a joint, possibly motionless domain
wall with alternating $q = \pm 1$ segments.  In fact, the steady
states which a randomly prepared system approaches with time take the
form of such piecewise domain walls containing HQVs.  Ordered yet
nonstationary states are also possible.  For instance, two curvilinear
$q = \pm 1$ segments can form a closed contour encapsulating a spot of
one domain inside the other which runs as a composite soliton.
Alternatively, two half-open domain walls rotate around a HQV pair and
so produce rotation of domains in a fixed environment.  All such
phenomena imply that the $q = \pm 1$ domain walls balance each other
dynamically on a large scale.

The paper is organized as follows.  In Sec.~\ref{sec:symmetry}, we
discuss the very possibility of a free evolution of coherently driven
BECs.  Section~\ref{sec:vacuum} describes the vacuum states.  Having
them defined, in Sec.~\ref{sec:1d}, we formulate a boundary-value
problem for domain walls and analyze its solutions in the
one-dimensional case.  Section~\ref{sec:2d} deals with domain walls
and HQV molecules in two dimensions, paying special attention to
long-range collective states (\ref{sec:2d:synthetic}) and spontaneous
phenomena (\ref{sec:2d:spontaneous}).  Section~\ref{sec:conclusion}
contains concluding remarks.  In Appendix~\ref{sec:app:parameters}, we
specifically consider the system of cavity polaritons and its
parameters.  Appendix~\ref{sec:app:vacuum} contains the exact solution
to the problem of the vacuum states and respective elementary
excitations.  The work also has four Supplemental Materials with video
presentations of dynamics in some noteworthy cases.

\section{Symmetry considerations}
\label{sec:symmetry}

In this introductory section, we briefly explain why a driven BEC can
behave ``like a freely evolving system''.  That is a bit arguable; one
might suggest that, conversely, the driving field governs both
components of the spinor order parameter, in analogy to a damped
oscillator driven by a harmonic force, which prevents formation of
vortices.  It turns out, however, that such a forced evolution is not
absolutely necessary.

Substituting $\psi_\pm = \sqrt{n_\pm} e^{i \phi_\pm}$ in
\eqref{eq:mean-field} leads one (\cite{Pitaevskii.book}) to the
following continuity equation,
\begin{multline}
  \label{eq:continuity}
  \frac{\partial}{\partial t} \left( n_+ + n_- \right) =
  -\nabla \left( \mathbf{j}_+ +\mathbf{j}_- \right) -
  \frac{2 \gamma}{\hbar} \left( n_+ + n_- \right) \\ {} -
  \frac{2 f}{\hbar} \left(
    \sqrt{n_+} \sin \phi_+ + \sqrt{n_-} \sin \phi_-
  \right),
\end{multline}
where $\mathbf{j}_\pm$ are currents, defined as
\begin{equation}
  \label{eq:continuity:currents}
  \mathbf{j}_\pm = -\frac{i \hbar}{2 m}
  \left(
    \psi_\pm^* \nabla \psi_\pm^\vphast -
    \psi_\pm^\vphast \nabla \psi_\pm^*
  \right)
  = \frac{n_\pm \hbar}{m} \nabla \phi_\pm.
\end{equation}
The last term in \eqref{eq:continuity} represents the source of
excitation.  In the case of a \emph{one-component} system (i.\,e.,
when one of $\psi_\pm$ is zero or $\psi_+$ and $\psi_-$ are
identical), this driving term invalidates the conventional
steady-state vortex solutions such that $\nabla n \neq 0$ and
$\nabla \phi \neq 0$ but
$(\nabla n) (\nabla \phi) = n \nabla^2 \phi = 0$; moreover, all actual
steady states are uniform as long as the driving field has zero wave
number~\cite{Gavrilov2020.usp.ru.en}.  We will show that in the
\emph{two-component} system, by contrast, domain walls and HQV
molecules are not only admitted, but arise spontaneously.  Domain
walls move in space ``freely'' in the sense that the driving
coefficient,
\begin{equation}
  \label{eq:K}
  K = -\sqrt{n_+} \sin \phi_+ - \sqrt{n_-} \sin \phi_-,
\end{equation}
is zero for each $\br$ for the $q = 0$ domain walls and has zero mean
for the $q = \pm 1$ ones, provided that $\gamma \to 0$.

The condition $K = 0$ is equivalent to
$\psi_+^\vphast + \psi_-^\vphast = \psi_+^* + \psi_-^*$ (which
represents a sort of weak symmetry) and also equivalent to invariance
of the density of energy $(\mathcal{H})$ under the transformation
$\psi_\pm \mapsto \psi_\pm e^{i \chi}$ for small $\chi$.  Directly
from \eqref{eq:energy} one can see that
$\mathcal{H} \bigl( \psi_\pm e^{i \chi} \bigr) - \mathcal{H} \bigl(
\psi_\pm \bigr)$ tends to $2 \chi f K$ when $\chi \to 0$.  In fact, it
is just a special case of Noether's theorem that invariance under
\emph{infinitesimal global phase shift} yields the continuity equation
with no external sources (e.\,g.,~\cite{Rubakov.book.CTGF.en}).
Therefore, it comes as no surprise that the $q = 0$ states with
$K = 0$ behave similarly to the usual phase domain walls which give
birth to HQV molecules via the transverse instability.  The symmetry
of the $q = \pm 1$ states is reduced further, since $K(\br)$ becomes
nonzero yet antisymmetric across the wall, which manifests broken
parity and involves a row of uncommon features.  Nevertheless, when
seen as individual objects (on such a scale that their fine structure
is indistinguishable), these domain walls also demonstrate an
effectively sourceless solitonic motion.

The above considerations help us to compare our results to the
previously known mechanisms of vortex formation in resonantly excited
polariton systems.  In all cases, the key point is to get rid of the
phase locking of the driven fluid with respect to the driving optical
field.  Earlier, this was done by making the system explicitly
inhomogeneous in space or time.  For instance, vortices can be
``injected'' directly by means of short-term excitation pulses with
appropriate shapes~\cite{Boulier2015, Dominici2015, Dominici2018}.
Alternatively, one can excite the microcavity by a focused laser beam
having a nonzero in-plane momentum, so that pumped polaritons run out
of the excitation spot and then scatter on artificially introduced
impurities~\cite{Carusotto2013, Pigeon2011, Amo2011, Pigeon2017}.  The
third mechanism employs the scattering of the pumped mode into a pair
of new macroscopically occupied modes that have different
frequencies~\cite{Whittaker2007, Krizhanovskii2010, Marchetti2010,
  Sanvitto2010, Dagvadorj2015}.  In all such situations (see
also~\cite{Koniakhin2019, Claude2020}), some part of a driven system
gets ``unlocked'' at a specific place and exhibits topological defects
only there.  By contrast, our scenario does not rely upon any
inhomogeneity of the model and $f$ is constant, so that domain walls
and vortices originate spontaneously and move freely.  Relevant
parameters of the polariton system are discussed in
Appendix~\ref{sec:app:parameters}.

\section{Vacuum states}
\label{sec:vacuum}

\begin{figure}
  \centering
  \includegraphics[width=\linewidth]{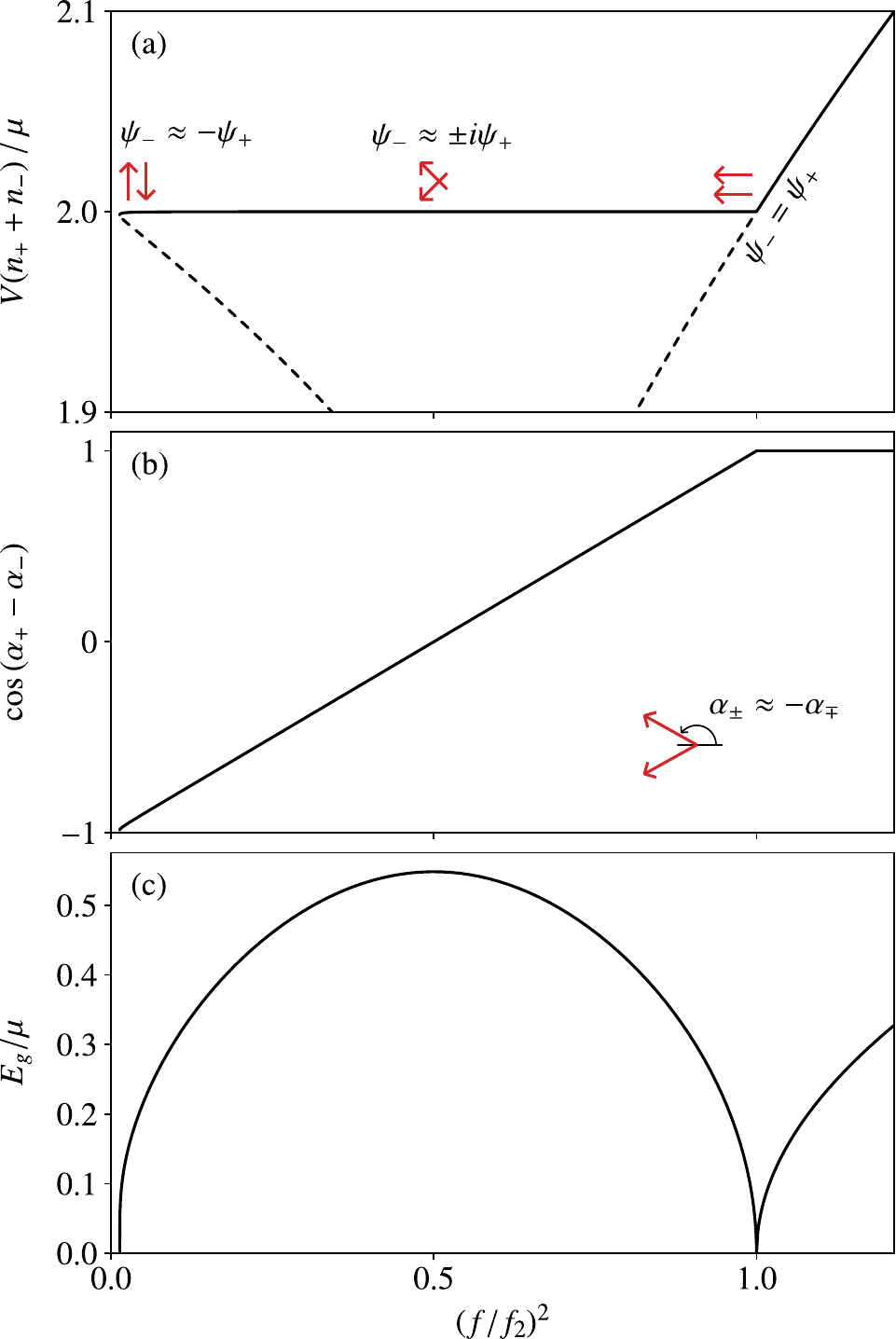}
  \caption{\label{fig:vac} Vacuum states depending on $f^2$.
    (a)~Density of the full number of particles according to
    \eqref{eq:vac:asymm:f} for $f < f_2$ and \eqref{eq:vac:symm} for
    $f \ge f_2$.  The dashed curves show unstable solutions.
    (b)~Cosine of the relative phase of $\psi_\pm$ according to
    \eqref{eq:vac:asymm:phase-diff} (exactly) or
    \eqref{eq:asymm:phase} (approximately).  The arrows in (a) and (b)
    indicate individual phases $\alpha_\pm$.  (c)~Energy gap between
    the vacuum state and elementary excitations according to
    \eqref{eq:exc:asymm} for $f < f_2$ and \eqref{eq:exc:symm} for
    $f \ge f_2$ (exactly) or \eqref{eq:exc:asymm:gap} for $f < f_2$
    (approximately).  Parameters: $D = \Omega$,
    $\gamma / \Omega = 5 \times 10^{-3}$.}
\end{figure}

The vacuum states, which are purely one-mode solutions to
Eqs.~\eqref{eq:mean-field}, exist for each~$f$, because the nonlinear
interaction is defocusing $(V > 0)$ and the model is spatially
uniform; and since the system is dissipative $(\gamma > 0)$, both
order parameters $\psi_\pm$ can be expected to oscillate at the
frequency of the driving field (taken as zero).  However, although the
model is explicitly spin-symmetric, the one-mode solutions with
$\psi_+ = \psi_-$ are unstable a wide range of~$f$.

In Appendix~\ref{sec:app:vacuum}, we find all vacuum states and
elementary excitations in a general form.  Here, we only illustrate
the main features of the relevant solutions at
$\gamma \ll D \sim \Omega$.

The considered vacuum states possess a reduced spin symmetry of the
form $\psi_+^* \approx \psi_-^\vphast$.  They exist in the interval of
$V f^2$ from $V f_1^2 = 2 \gamma \Omega (D + \Omega)$ to
$V f_2^2 = \Omega^2 (D + \Omega / 2)$ and, moreover, turn out to be
the sole kind of stable one-mode solutions if
$f_1^2 \le f^2 \lesssim 0.3 f_2^2$.  Densities $n_\pm = |\psi_\pm|^2$
are almost independent of $f$ and equal to $\mu / V$, where
\begin{equation}
  \label{eq:mu}
  \mu = D + \frac{\Omega}{2}.
\end{equation}
The phases of spin components in two vacuum states,
$\alpha_{\pm(\mp)}$, are nearly equal to $\pm \alpha$, where $\alpha$
varies from about $\pi/2$ to $\pi$ as
$\alpha \approx \arccos(-f / f_2)$, provided that $f > 0$.  At
$f = f_2$, the solution becomes symmetric and single-valued.
Characteristic dependences of $n_+ + n_-$ and
$\cos(\alpha_+ - \alpha_-) \approx \cos 2 \alpha$ on $f^2$ are shown
in Figs.~\ref{fig:vac}(a) and~\ref{fig:vac}(b).

Bearing in mind optical experiments, it is convenient to express the
solutions in terms of the Stokes vector components.  Proceeding from
unitary transformation $\psi_\pm = (\psi_H \mp i \psi_V) / \sqrt{2}$,
where $\psi_{H,V}$ are the fields with mutually orthogonal
``horizontal'' and ``vertical'' polarizations, one defines
\begin{equation}
  \label{eq:stokes12}
  S_1 = \frac{|\psi_H|^2 - |\psi_V|^2}{|\psi_H|^2 + |\psi_V|^2},
  \quad
  S_2 = \frac{\psi_H^* \psi_V^\vphast + \psi_V^* \psi_H^\vphast}
  {|\psi_H|^2 + |\psi_V|^2},
\end{equation}
which are the degrees of linear polarization in two Cartesian bases
rotated relative to each other by $\pi / 4$.  The third component of
the Stokes vector,
\begin{equation}
  \label{eq:stokes3}
  S_3 = \frac{|\psi_+|^2 - |\psi_-|^2}{|\psi_+|^2 + |\psi_-|^2},
\end{equation}
represents the degree of circular polarization.  The whole Stokes
vector $\mathbf{S} = (S_1, S_2, S_3)$ has the length of unity.

Since $n_+ \approx n_-$ and, thus, $S_3 \approx 0$ for
$f \in (f_1, f_2)$, the field is polarized almost linearly.
Specifically,
\begin{equation}
  \label{eq:lin-pol}
  \dbinom{\psi_H}{\psi_V} =
  \sqrt{2 \mu} \,
  \dbinom{\cos \alpha}{\pm \sin \alpha},
  \quad S_1 = \cos 2 \alpha,
  \quad S_2 = \pm \sin 2 \alpha.
\end{equation}
Thus, $\alpha$ equals the angle between the polarization directions of
the driven and driving fields, $\psi_{H,V}$ and
$f_{H,V} = (\sqrt{2} f, \, 0)$.  The two states forming the phase
doublet at a given $f$ differ in the sign of $S_2$ that is directly
measurable.

\begin{figure}
  \centering
  \includegraphics[width=\linewidth]{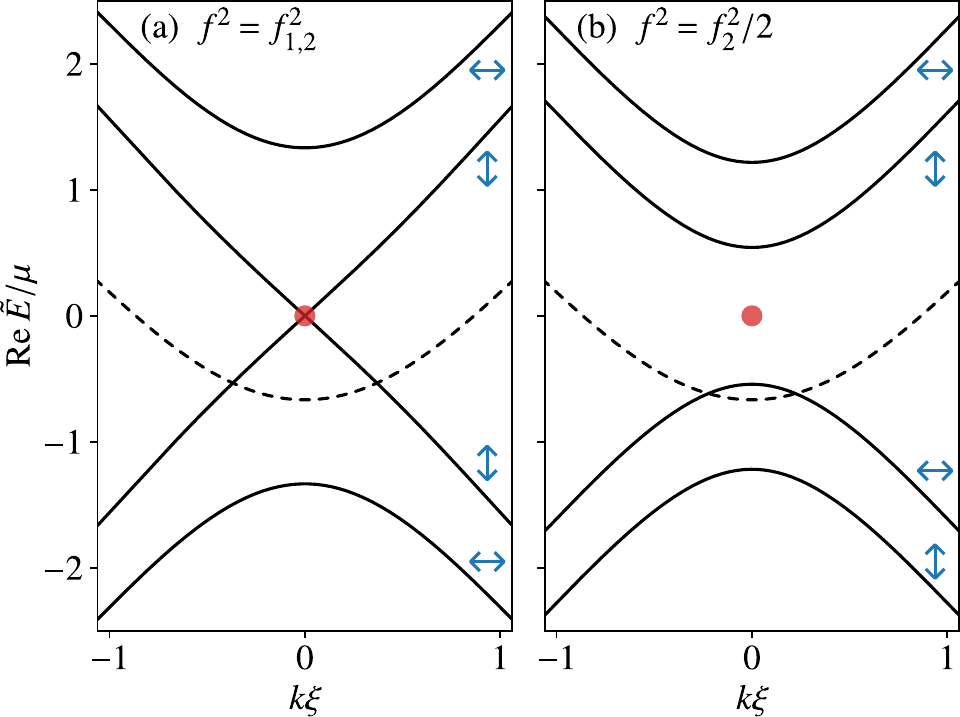}
  \caption{\label{fig:vac:excitations} Spectra of elementary
    excitations depending on wave number $k$ at $f = f_1$ or $f = f_2$
    (a) and $f = \sqrt{2} f_2$ (b) according to \eqref{eq:exc:asymm}
    (exactly) or \eqref{eq:exc:top} (approximately).  The dashed curve
    shows the bare dispersion law; the central dot is the driven mode.
    The gapless spectrum in (a) obeys the Bogoliubov formula
    \eqref{eq:exc:bgl}.  The arrows indicate polarization directions
    of excitations, horizontal $(\leftrightarrow)$ or vertical
    $(\updownarrow)$.  Parameters as in Fig.~\ref{fig:vac};
    $\xi = \hbar / \sqrt{2 m \mu}$.}
\end{figure}

The asymptotic stability of the solutions is ensured by the gap
between the vacuum state and elementary excitations.  Its width $E_g$
as a function of $f$ is shown in Fig.~\ref{fig:vac}(c).  When $E_g$
turns to zero at $f = f_{1,2}$, the dispersion law takes the
Bogoliubov form~\eqref{eq:exc:bgl}.  Remarkably, in this case quantity
\eqref{eq:mu} plays the role of an equilibrium chemical potential.
Typical spectra of excitations, gapped and gapless, are shown in
Fig.~\ref{fig:vac:excitations}.

Starting with Figs.~\ref{fig:vac} and \ref{fig:vac:excitations}, in
all illustrations we use a system of ``natural'' physical units.
Specifically, $\mu$ represents the densities of the numbers of
particles $(V n_\pm)$ and typical shifts of energy levels; the
respective healing length $\xi = \hbar / \sqrt{2 m \mu}$ is the
natural scale and $\xi^{-1}$ is the wave number.  The sound velocity
in a conventional BEC, $v_s = \sqrt{\mu / m}$, serves as a
characteristic velocity.

To simplify the discussion, most of the time we will imply the
approximation~$\gamma \ll \Omega \sim D$.  Let us indicate, however,
what changes beyond its scope.  In accordance with the exact solutions
derived in Appendix~\ref{sec:app:vacuum}, nonzero $\gamma$ necessarily
means nonzero $n_+ - n_-$ and $\alpha_+ + \alpha_-$ (each of these
quantities differs in sign for the two vacuum states).  Besides,
increasing $\gamma$ leads to the increase of the threshold
$f_1^2$~\eqref{eq:asymm:f1} at which the considered vacuum states
appear.  On the other hand, a significant imbalance of $\Omega$ and
$D$ can produce qualitatively different branches of solutions with
high~$|S_3|$ in a range of small~$f^2$, in which case phase domain
walls cease to exist.

\section{Domain walls in one dimension}
\label{sec:1d}

\begin{figure}[!b]
  \centering
  \includegraphics[width=\linewidth]{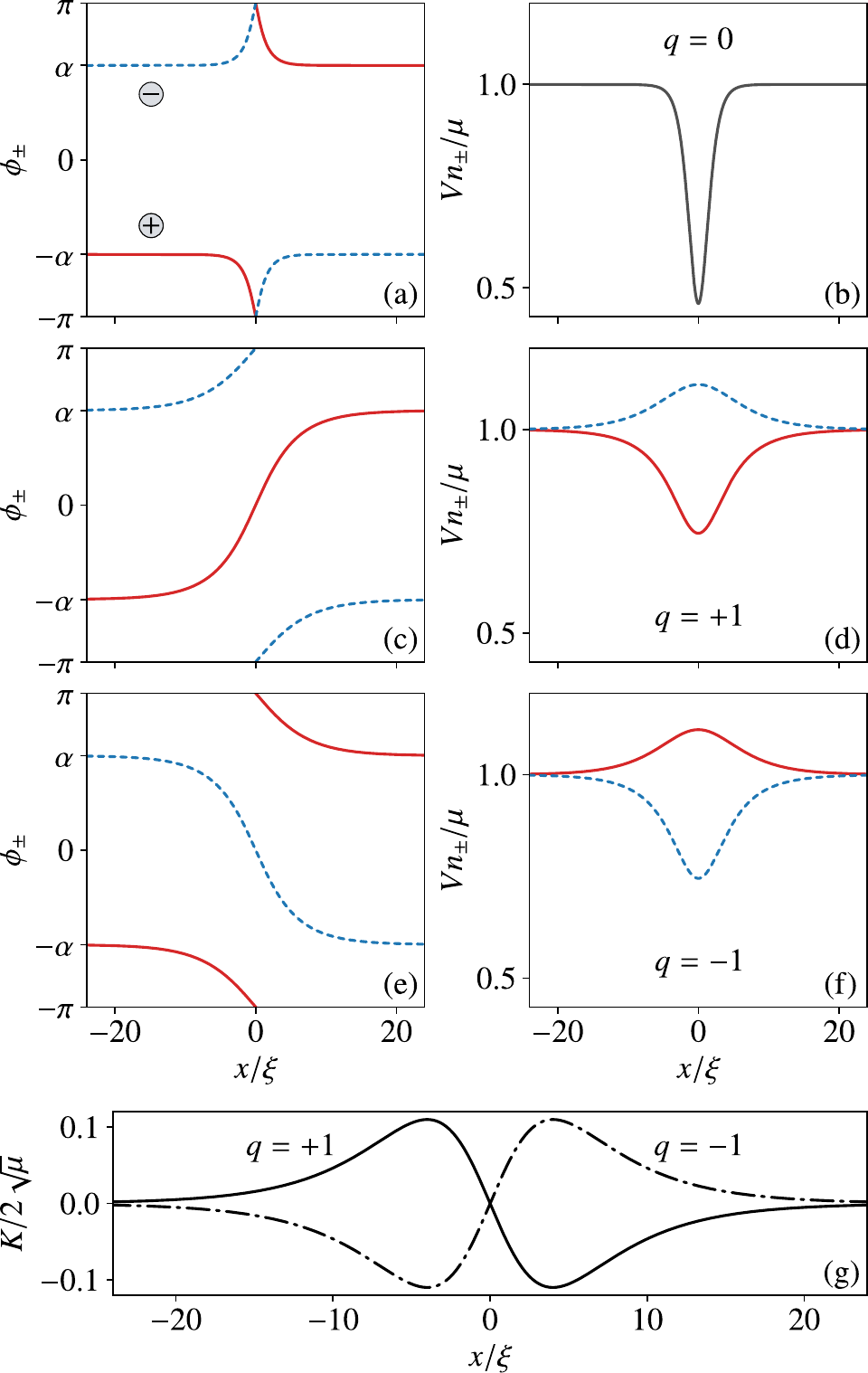}
  \caption{\label{fig:bvp:types} (a)--(f) Three solutions to
    boundary-value problem \eqref{eq:soliton}--\eqref{eq:bc2} for
    $v = 0$ and $f^2 / f_2^2 = 0.1$.  The left and right sides
    represent $\phi_\pm = \arg \psi_\pm$ and $n_\pm = |\psi_\pm|^2$,
    respectively; solid and dashed curves correspond to individual
    components ($+$ and $-$).  (g)~Quantity $K$~\eqref{eq:K} across
    the $q = \pm 1$ domain walls.  Parameters: $D = \Omega$,
    $\gamma / \Omega = 10^{-6}$.}
\end{figure}

Let us find the shape of domain walls in a relatively simple case of
the one-dimensional system.  Substituting
$\psi_\pm(x, t) = \psi_\pm(x - vt)$ into Eqs.~\eqref{eq:mean-field}
leads one to the following system of time-independent equations in the
soliton frame of reference,
\begin{equation}
  \label{eq:soliton}
  \frac{\hbar^2}{2m} \frac{d^2 \psi_\pm}{d x^2}
  - i \hbar v \frac{d \psi_\pm}{d x}
  + \left( D + i \gamma - V \psi_\pm^* \psi_\pm^\vphast \right)
  \psi_\pm^\vphast
  - \frac{\Omega}{2} \psi_\mp^\vphast
  -f = 0.
\end{equation}
The boundary conditions correspond to the vacuum states and,
therefore, should be set rigidly, e.\,g.,
\begin{gather}
  \label{eq:bc1}
  \psi_-(-\infty) = \psi_+(+\infty) = \sqrt{n_1} e^{i \alpha_1},\\
  \label{eq:bc2}
  \psi_+(-\infty) = \psi_-(+\infty) = \sqrt{n_2} e^{i \alpha_2},
\end{gather}
where $n_{1,2}$ and $e^{i \alpha_{1,2}}$ are determined by
\eqref{eq:vac:asymm:u12}--\eqref{eq:vac:asymm:phase-exp}.  Below, we
take very small $\gamma$ and assume that $n_1 = n_2$ and
$\alpha_1 = -\alpha_2 = \alpha > 0$.  Swapping the two domains,
i.\,e., choosing the opposite sign of $\alpha$ or exchanging the
right-hand sides in \eqref{eq:bc1} and \eqref{eq:bc2}, would result in
a physically equivalent situation.

In what follows, we solve this boundary-value problem numerically by
means of the fourth-order collocation algorithm~\cite{Kierzenka2001}
implemented in SciPy~\cite{Virtanen2020}.  Technically, to do so one
has to rewrite \eqref{eq:soliton} in the form of eight real-valued
equations of the first order.

\begin{figure}[!b]
  \centering
  \includegraphics[width=\linewidth]{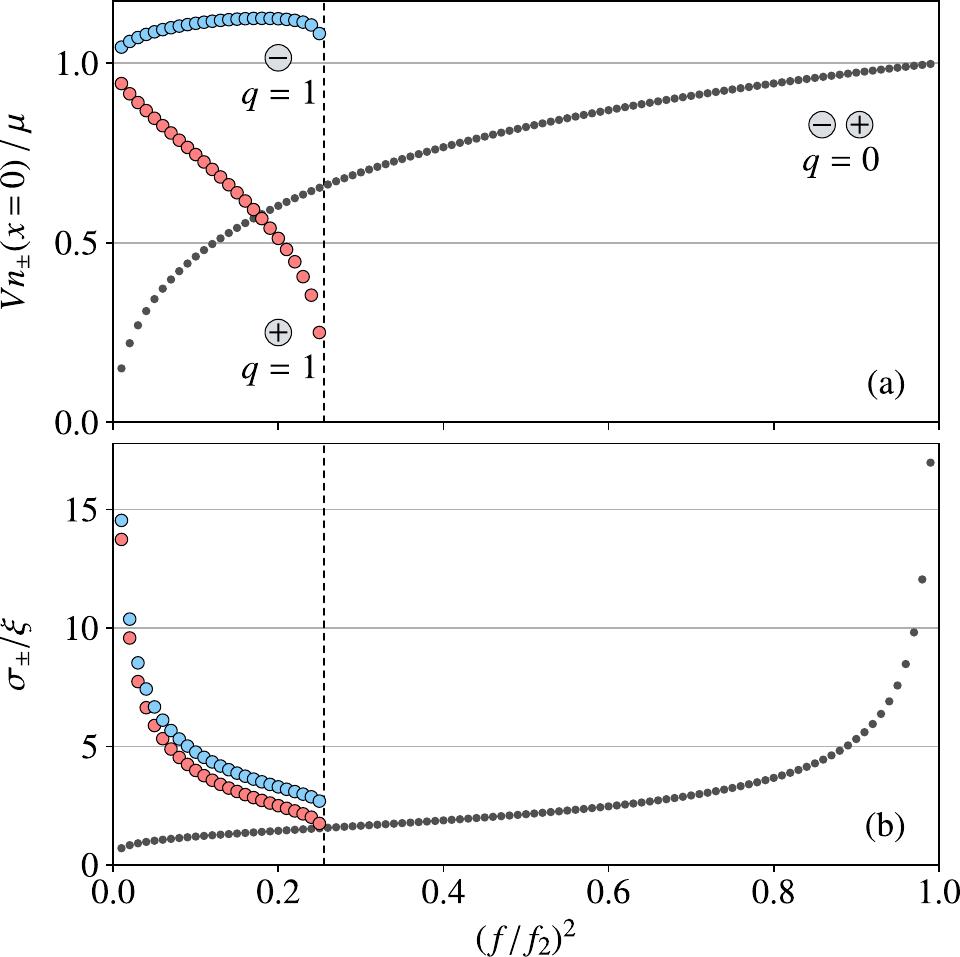}
  \caption{\label{fig:bvp:series:pump} Peak values of $n_\pm$ (a) and
    spatial widths $\sigma_\pm$ (b) of motionless domain walls
    depending on $f^2$.  The dashed vertical line indicates the
    limiting value of $f^2$ at which the $q = \pm 1$ domain walls
    cease to exist.  The solution with $q = -1$ (not shown) is
    obtained by exchanging `$+$' and `$-$' for $q = 1$.  Parameters as
    in Fig.~\ref{fig:bvp:types}.}
\end{figure}

\begin{figure}[!b]
  \centering
  \includegraphics[width=\linewidth]{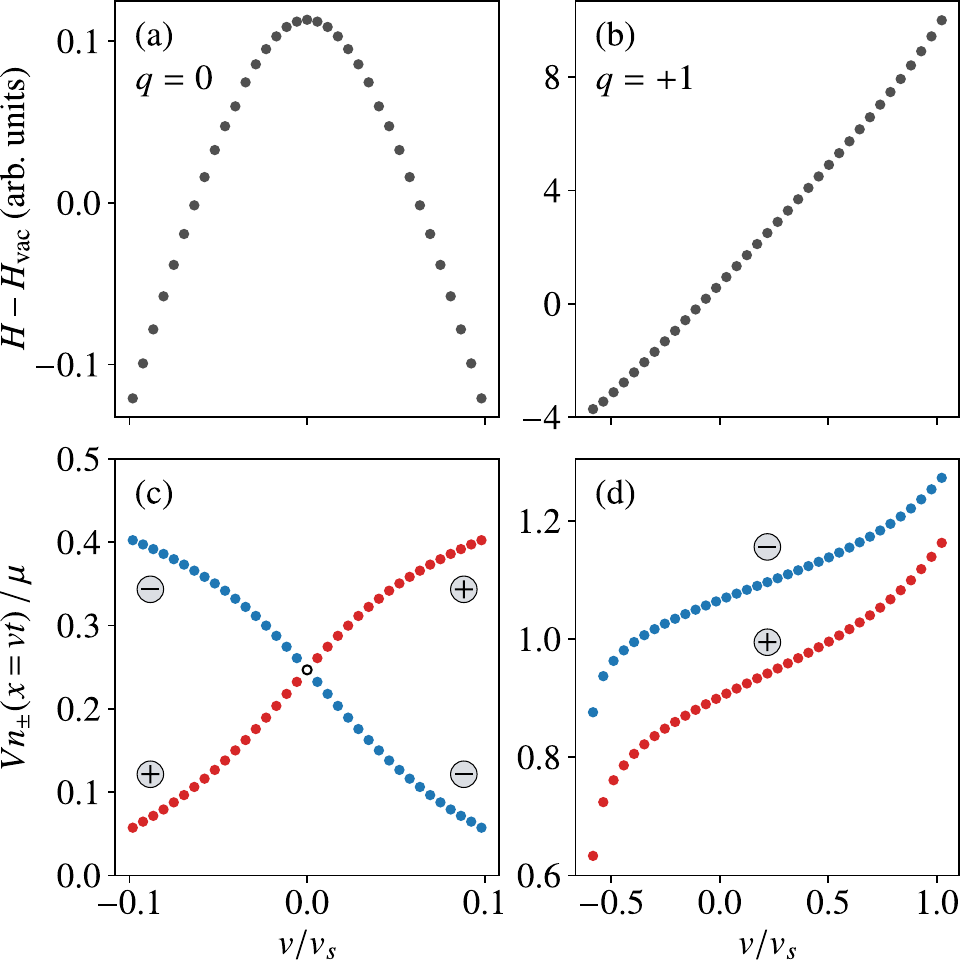}
  \caption{\label{fig:bvp:series:velocity} Energy~\eqref{eq:energy}
    (a, b) and peak values of $n_\pm$ (c, d) as functions of $v$ for
    the domain walls with $q = 0$ (a, c) and $+1$ (b, d) at
    $f^2 / f_2^2 = 0.025$.  The solution with $q = -1$ (not shown) is
    obtained by reversing time $(v \mapsto -v)$ and exchanging `$+$'
    and `$-$' for $q = 1$.  Parameters as in Fig.~\ref{fig:bvp:types};
    $H_\mathrm{vac}$ is the vacuum energy; $v_s = \sqrt{\mu / m}$.}
\end{figure}

\subsection{Static domain walls $(v = 0)$}
\label{sec:1d:static}

For each given set of parameters and boundary conditions, the above
problem can have several topologically different solutions which are
distinguished by the total phase variation
\begin{equation}
  \label{eq:q}
  q = \frac{1}{2 \pi} \int\limits_{-\infty}^{+\infty}
  \,
  \left(
    \frac{d \phi_+}{dx} + \frac{d \phi_-}{dx}
  \right) dx.
\end{equation}
For consistency, swapping the vacuum states in \eqref{eq:bc1} and
\eqref{eq:bc2} should be accompanied by the $x$ axis reversal
in~\eqref{eq:q}.  Typical solutions with $q = 0$ and $q = \pm 1$
obtained at $v = 0$ are represented in Fig.~\ref{fig:bvp:types}.

The $q = 0$ domain walls [Figs.~\ref{fig:bvp:types}(a) and
\ref{fig:bvp:types}(b)] retain the symmetry of the form
$\psi_+^* = \psi_-^\vphast$ for each $x$.  Generally, if
$\psi_+^* = \psi_-^\vphast$, phases $\phi_+$ and $\phi_-$ coincide at
the center of a domain wall $(x = 0)$, being equal to $0$ or
$\pm \pi$.  The solutions with $\phi_\pm(0) = 0$ never occurred in our
calculations; notice that they would have maximized the last term
in~\eqref{eq:energy} since $f > 0$.  In turn, the solutions with
$\phi_\pm(0) = \pm \pi$ exist in the whole interval $(f_1, f_2)$.

The $q = \pm 1$ domain walls
[Figs.~\ref{fig:bvp:types}(c)--\ref{fig:bvp:types}(f)] have nonzero
spin polarizations $S_3$ and co-directed currents
$\mathbf{j}_\pm \propto n_\pm \nabla \phi_\pm$ even at $v = 0$.
Moreover, $\nabla (\mathbf{j}_+ + \mathbf{j}_-)$ is also nonzero,
which can be attributed only to external driving.  To illustrate this
point, Fig.~\ref{fig:bvp:types}(g) shows function $K(x)$ that
determines the source term in the continuity
equation~\eqref{eq:continuity}.  $K(x) > 0$ means that the system has
an external source of particles at a given $x$, compensated by a
``negative source'' at~$-x$.  For $q = +1$, the whole picture suggests
that particles tunnel from positive to negative $x$ under the external
field.  In turn, on both sides of $x = 0$ these sources are
counterbalanced by the currents with
$\nabla (\mathbf{j}_+ + \mathbf{j}_-) \neq 0$, so that
Eq.~\eqref{eq:continuity} is satisfied statically
$(\partial_t n_\pm = 0)$.

Figure~\ref{fig:bvp:series:pump} shows how the spatial widths and peak
values of $n_\pm$ depend on $f$ for different types of domain walls.
The widths $\sigma_\pm$ are calculated by fitting the numerically
obtained phase patterns $\phi_\pm(x)$ with appropriately scaled kinks,
\begin{equation}
  \label{eq:kink}
  \phi(x) = \phi_1 + \frac{2}{\pi} \left( \phi_2 - \phi_1 \right)
  \arctan e^{x / \sigma},
\end{equation}
where $\phi_{1,2} = \phi(\mp \infty)$.  The $q = 0$ and $q = \pm 1$
domain walls are seen to behave oppositely.  The $q = 0$ solution has
the greatest dip in $n(x)$ as well as the minimum $\sigma$ at
$f = f_1$, becomes less pronounced with increasing $f$, and vanishes
at $f = f_2$.  Since the states with $\psi_+^* = \psi_-^\vphast$ are
polarized linearly, their features are better understood in terms of
the $\psi_{H,V}$ rather than $\psi_\pm$ fields.  The key point is the
phase reversal of $\psi_V$ between the vacuum states
[Eq.~\eqref{eq:lin-pol}], owing to which the $q = 0$ domain walls can
be seen as dark solitons in~$\psi_V$.  With increasing $f$, they
become less pronounced, because the fraction of $\psi_V$ is lowered.
The width $\sigma = \hbar / \sqrt{2 m E}$ of such domain walls roughly
corresponds to energy $E = V \psi_V^2$ in the vacuum states.

In turn, the width of the $q = \pm 1$ domain walls corresponds to
their spin imbalance $(E \sim V \langle | n_+ - n_- | \rangle)$ which
grows with~$f$.  Notice that the phase jumps of two components across
a domain wall, $\pm 2 \alpha$ and $\pm 2 \pi \mp 2 \alpha$, become
essentially different at high $f$ and the greater one $(\pm 2 \alpha)$
is accompanied by a growing dip in $n(x)$.  As a result, the spin
polarization of the currents remains small $(j_+ \approx j_-)$ for
each~$f$ in spite of increasing $|S_3|$.  This branch of solutions is
seen to terminate at some point.

\subsection{Motion of domain walls}
\label{sec:1d:moving}

Let us find out how the characteristics of domain walls depend on
their state of motion.  In atomic BECs, the energy of a moving soliton
typically has a quadratic dependence on its velocity,
$ H(v) = H(0) + m^\star v^2 / 2$, which allows one to introduce the
soliton mass $m^\star$~\cite{Kamchatnov2008,Gallemi2019}.  That mass
can be negative, in which case a \emph{dark} soliton gets accelerated
upon losing energy through its interaction with thermal excitations.

Figure~\ref{fig:bvp:series:velocity} shows the dependences of $H$ and
peak $n_\pm$ on $v$ for the $q = 0$ and $q = 1$ domain walls slightly
above the $f_1$ point.  The $q = 0$ state behaves similarly to the
$2 \pi$ magnetic solitons in atomic BECs~\cite{Qu2017,Yu2021,Yu2022}.
It has a negative mass and nonzero $S_3$ whose sign depends on the
direction of motion.  A pair of the $q = 0$ states with opposite
velocities transform into each other continuously.  The greatest $|v|$
at which this kind of solutions exists is comparable to $0.1 v_s$.  As
a rule, the sign of the soliton mass is reversed with increasing~$f$.

By contrast, the $q = \pm 1$ domain walls are not characterized by any
mass in the usual sense.  They have certain signs of $S_3$
irrespective of~$v$ and tend to move in certain preferred directions.
The energy is lowered when $v$ and $q$ have opposite signs.
Remarkably, both $n_+$ and $n_-$ decrease upon lowering $H$, so that a
\emph{gray} soliton becomes more \emph{dark} in the course of its
spontaneous acceleration.  In this respect, our system behaves
conversely to the usual BECs in which increasing $v$ acts to smoothen
a dark soliton until it reaches the sound velocity and turns into a
plane wave.  Notice as well that the minimum $H$ in
Fig.~\ref{fig:bvp:series:velocity}(b) is less than the vacuum energy.
Therefore, a domain wall moving with its greatest negative velocity
$v \lesssim -0.5 v_s$ is expected to be energetically stable.

The above predictions have been verified numerically by solving the
dynamical equations~\eqref{eq:GP} for a two-dimensional polariton
system of size $500 \, \xi \times 30 \, \xi$.  A video presentation of
the solution can be found in Supplemental Material~I~\footnote{See
  \href
  {https://arxiv.org/src/2505.09553v4/anc/Supplemental_Material_I.mp4}
  {Supplemental Material~I} for a video presentation of the soliton
  motion.}.  In the long dimension (axis~$x$), the boundary conditions
were zero owing to high potential walls at both ends, whereas in the
short dimension~($y$) they were periodic.  For each $y$, the initial
state along the $x$ axis was the $q = 1$ domain wall obeying
Eqs.~\eqref{eq:soliton}--\eqref{eq:bc2} at $D = \Omega$,
$\gamma / \Omega = 1.5 \times 10^{-4}$, and $f_2 / f_2^2 = 0.025$.
Finite fluctuations came from the fact that the chosen initial state
is not an exact solution for the system in a box.  The dynamical
equations were solved using the adaptive RK5(4) method by Dormand and
Prince~\cite{Dormand1980}.  In agreement with our qualitative
analysis, it turned out that interaction with fluctuations makes the
soliton run in the opposite direction relative to~$\nabla \phi_\pm$.
With time, this soliton becomes ``gray'' in both spin components,
reaches its greatest velocity, and then runs steadily until it hits
the end wall.  Here, the soliton gets reflected, changes the sign of
$q$, and starts moving back towards the other end wall.  Thus, the
overall dynamics takes the form of periodic spatiotemporal
oscillations of $\phi_\pm$ and~$S_2$.

Up to now, we supposed that the decay rate $\gamma$ is extremely
small.  When it is increased, everything becomes more complicated, in
particular, the solutions to Eqs.~\eqref{eq:mean-field} no longer
satisfy the solitonic Eqs.~\eqref{eq:soliton}.  (Technically, the
minimized left-hand sides of \eqref{eq:soliton} tend to certain small
yet nonzero residual values that increase with~$\gamma$.)  On the
other hand, the asymptotic stability of the phase domains \textit{per
  se} is only enhanced, because all excitations inside them tend to
decay on the time scale of $\hbar / \gamma$ [Eq.~\eqref{eq:exc:top}].
As a result, the domain walls survive and still have the $q = \pm 1$
types at small~$f$, however, they become nonstatic even in the soliton
frame of reference.  The above-considered numerical experiment also
displays fast and irregular oscillations of $n_\pm(x \, {=} \, vt)$,
which suggests that the persistence of the solitonic interface should
rather be ascribed to the internally stable domains.  In this respect
the discussed system noticeably differs from the freely evolving
atomic BECs in which the (in)stabilities of domains and domain walls
mean the same thing.

Finally, let us notice that the solutions to Eqs.~\eqref{eq:soliton}
at $v > 0$ and $\gamma \to 0$ can also converge to higher-order states
with $q \ge 2$.  Under normal conditions, such states are unstable in
two dimensions, so we do not analyze them in the current work.  The
greatest positive velocity at which the solutions robustly converge to
the $q = 1$ states in Figs.~\ref{fig:bvp:series:velocity}(b),
\ref{fig:bvp:series:velocity}(d) is close to $v_s$, where $n_\pm$
start growing rapidly.  The greatest negative velocity, at which
$n_\pm$ abruptly drop down, has a smaller absolute value.

\begin{figure}[!b]
  \centering
  \includegraphics[width=\linewidth]{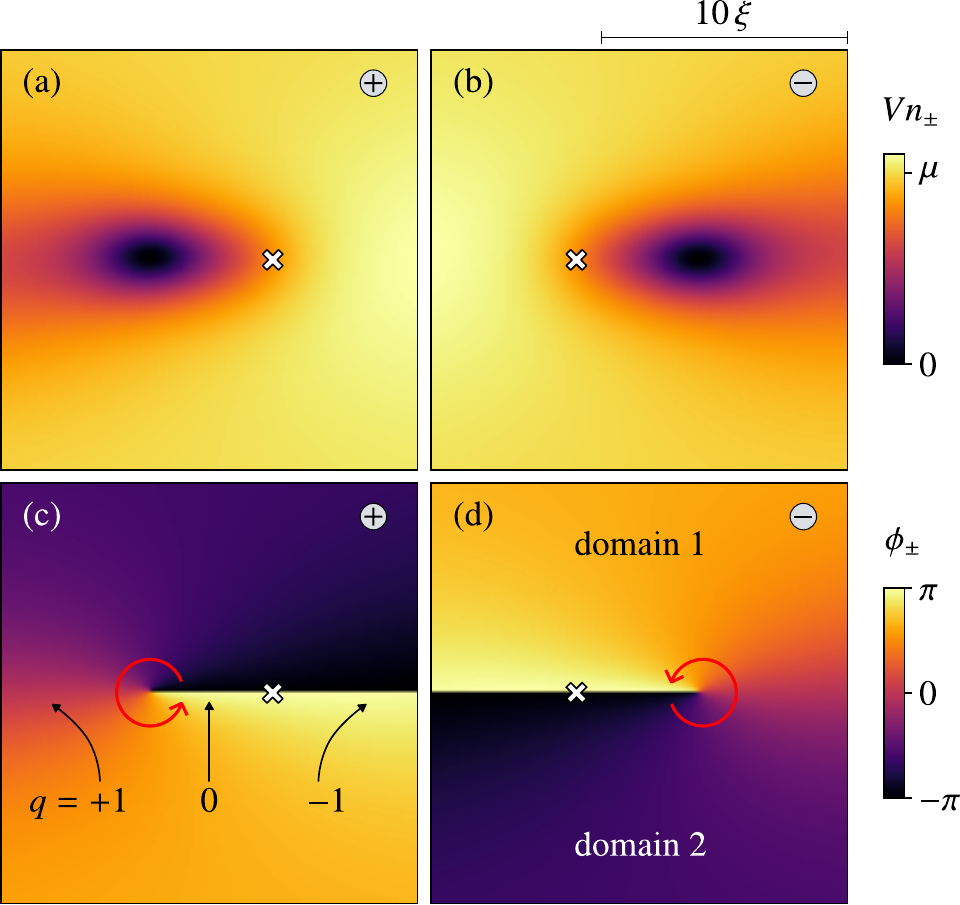}
  \caption{\label{fig:2d:shape} Characteristic solution having two
    phase domains, the $q = 0$ and $q = \pm 1$ domain walls, and a HQV
    molecule.  The top and bottom sides show $n_\pm$ and $\phi_\pm$;
    the left and right sides correspond to individual spin components
    ($+$ and $-$).  The cross marks $(\times)$ indicate positions of
    the opposite-spin vortices.  The circular arrows in (c) and (d)
    indicate the directions in which $\phi_\pm$ vary from $-\pi$
    to~$\pi$.  Parameters: $D = \Omega$,
    $\gamma / \Omega = 5 \times 10^{-3}$, $f^2 / f_2^2 \approx 0.13$.}
\end{figure}

\section{Domain walls in two dimensions}
\label{sec:2d}

We have found that the $q = \pm 1$ domain walls tend to move in space.
Here, we show that they naturally coexist and balance each other in a
two-dimensional system.  As a characteristic example,
Fig.~\ref{fig:2d:shape} shows a small piece of a domain wall with a
crossover between the $q = \pm 1$ states.  Two HQVs are seen to
connect them continuously, much as a domain wall itself connects phase
domains.  In the vicinities of both cores, one of the amplitudes
$|\psi_\pm| = \sqrt{n_\pm}$ is proportional to the respective
distance.  The occurrence of internally ordered HQV molecules in a
driven BEC agrees with the fact that quantity $K$~\eqref{eq:K} is
small and smooth even at the singular points.  In fact, $K$ is nonzero
only on both sides of the $q = \pm 1$ domain walls in accordance with
Fig.~\ref{fig:bvp:types}(g).  Zero mean value of $K$ suggests that the
whole pattern represented in Fig.~\ref{fig:2d:shape} can move freely.

\subsection{Synthetic examples}
\label{sec:2d:synthetic}

\begin{figure}
  \centering
  \includegraphics[width=\linewidth]{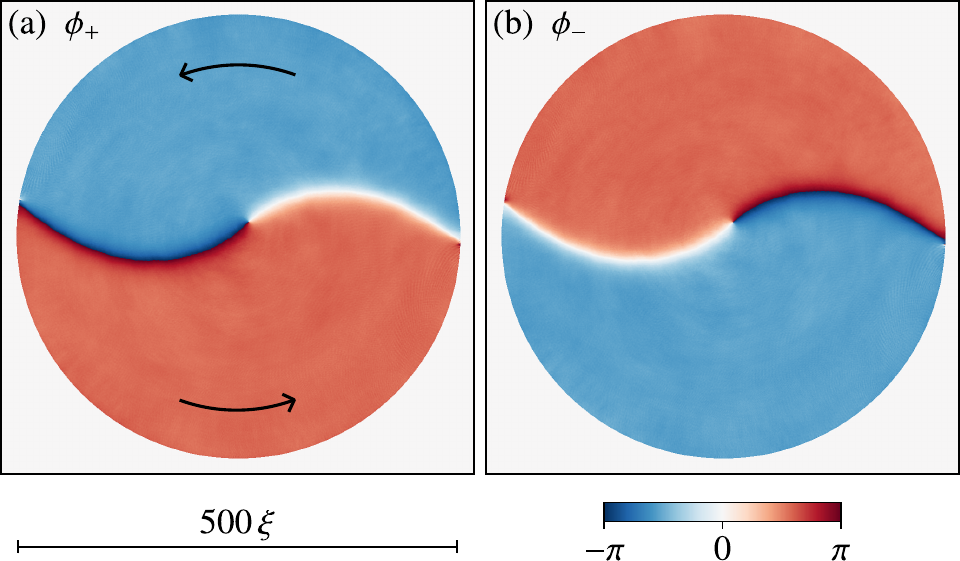}
  \caption{\label{fig:2d:disk} Persistently rotating solution with
    symmetric $q = \pm 1$ segments of a domain wall.  (a) and (b)
    represent phases $\phi_\pm$ of two spin components.  The arrows
    indicate the direction of rotation.  Parameters: $D = \Omega$,
    $\gamma / \Omega = 7.5 \times 10^{-4}$, $f^2 / f_2^2 = 0.025$.
    The dynamics is shown in Supplemental Material II~\cite{Note2}.}
\end{figure}

\begin{figure}
  \centering
  \includegraphics[width=\linewidth]{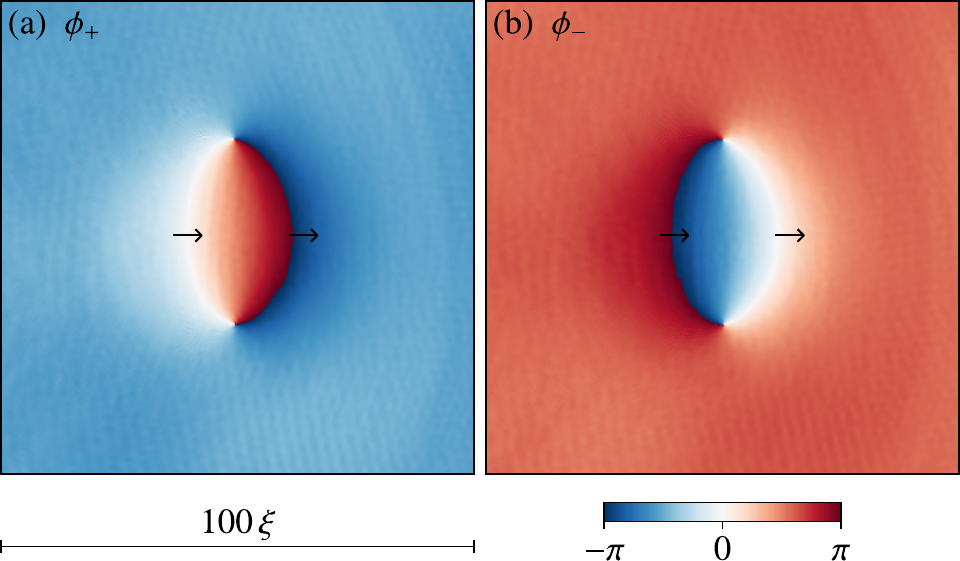}
  \caption{\label{fig:2d:drop} A closed contour of two $q = \pm 1$
    domain walls moving in space.  (a) and (b) represent phases
    $\phi_\pm$ of two spin components.  The arrows indicate the
    direction of motion.  Parameters as in Fig.~\ref{fig:2d:disk}.
    The dynamics is shown in Supplemental Material~III~\cite{Note3}.}
\end{figure}

\begin{figure}
  \centering
  \includegraphics[width=\linewidth]{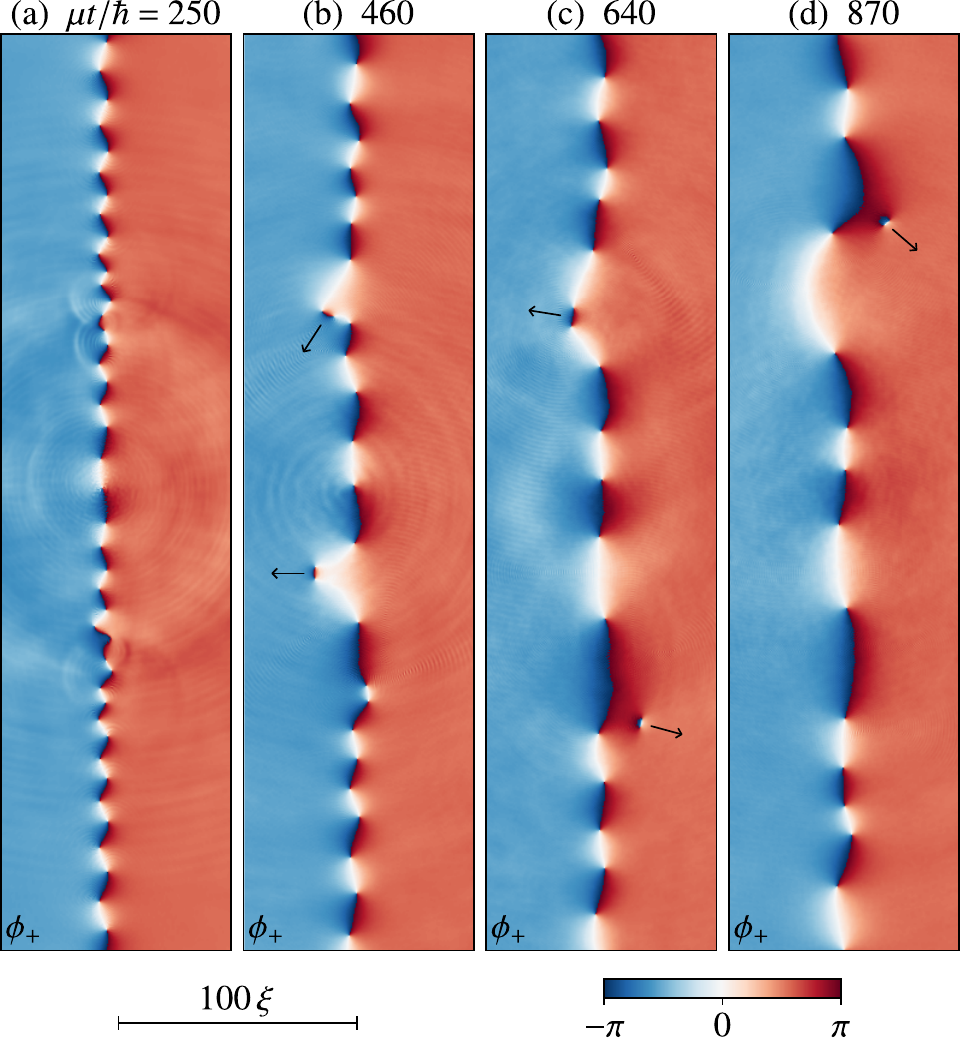}
  \caption{\label{fig:2d:chain} Sequence of spatially alternating
    $q = \pm 1$ states formed after the breakup of the $q = 0$ domain
    wall.  (a)--(d) correspond to different time moments.  The arrows
    indicate the directions of motion of small composite solitons.
    The color map represents~$\phi_+$.  Parameters as in
    Fig.~\ref{fig:2d:disk}.  The dynamics is shown in Supplemental
    Material~IV~\cite{Note4}.}
\end{figure}

Below, we demonstrate the main kinds of moving collective states with
the $q = \pm 1$ junctions.  We discuss three solutions to
Eqs.~\eqref{eq:GP} obtained on a $1024 \times 1024$ grid.  In all
cases, a high potential wall of radius $R \approx 250 \, \xi$ ensures
zero boundary conditions.  The initial states have two domains placed
at $x < 0$ and $x > 0$ and connected by domain walls
$\psi_\pm^{(q)}(x)$ whose type $q$ can be different depending on~$y$.
The driving amplitude~$f$ corresponds to
Fig.~\ref{fig:bvp:series:velocity} $(f^2 / f_2^2 = 0.025)$.  The
system dynamics is explicitly shown in the respective Supplemental
Materials II~\footnote{See \href
  {https://arxiv.org/src/2505.09553v4/anc/Supplemental_Material_II.mp4}
  {Supplemental Material~II} for a video presentation corresponding to
  Fig.~\ref{fig:2d:disk}.}, III~\footnote{See \href
  {https://arxiv.org/src/2505.09553v4/anc/Supplemental_Material_III.mp4}
  {Supplemental Material~III} for a video presentation corresponding
  to Fig.~\ref{fig:2d:drop}.}, and IV~\footnote{See \href
  {https://arxiv.org/src/2505.09553v4/anc/Supplemental_Material_IV.mp4}
  {Supplemental Material~IV} for a video presentation corresponding to
  Fig.~\ref{fig:2d:chain}.} that contain video presentations.

The first example has such initial state that $q = +1$ at $y \ge 0$
and $q = -1$ at $y < 0$.  The solution shows that a HQV molecule
rapidly appears near $\br = 0$, after which the whole picture becomes
self-consistent and evolves steadily.  As expected, the $q = \pm 1$
parts of the domain wall move in opposite directions and, since they
are connected, rotate around $\br = 0$.  A snapshot of $\phi_\pm(\br)$
at some time moment is shown in Fig.~\ref{fig:2d:disk}.  Clearly,
rotation implies that linear velocity $v$ grows with~$r$.  According
to Fig.~\ref{fig:bvp:series:velocity}(b), multiple solutions with
different $v$ indeed coexist for the same set of boundary conditions.
Since the range of $v$ is limited, the uniformly rotating patterns
cannot persist against arbitrarily increasing~$R$; it is remarkable,
however, that they are still possible for $R = 250 \, \xi$.  The
stability of the solution is related to the fact that its energy is
less than the vacuum one.  At the same time, a displacement of the HQV
molecule from the center eventually makes the whole system lose its
dynamical balance.  Analogous patterns often form spontaneously in
randomly prepared systems, yet, they usually have smaller sizes and
lifetimes.

In the second example, the system evolves from the initial state with
$q = +1$ at $|y| \le a$ and $q = -1$ at $|y| > a$, where $a$ is a
relatively small distance equal to $12.5 \, \xi = 0.05 \, R$.  The
outer segments have the same $q$ and rotate in opposite directions, so
they eventually meet and cancel each other everywhere except the
$|y| \lesssim a$ region in which the solution takes the form shown in
Fig.~\ref{fig:2d:drop}.  It is seen to be composed of two HQV
molecules and two $q = \pm 1$ domain walls that encapsulate some piece
of one domain inside the other.  In spite of opposite $q$, the average
phase gradients of the domain walls have the same direction, because
the respective pairs of their ``left'' and ``right'' domains are also
mutually opposite.  As a result, we have a composite soliton that
moves as a unified object preserving its shape.  Direct calculation
shows that energy~\eqref{eq:energy} is now greater than the vacuum
energy.  Thus, being stable in the dynamical sense, this pattern
suffers the Landau damping and gradually decreases in size until two
HQV molecules annihilate.  Noteworthy, it moves in the direction of
growing rather than decreasing phases, which corresponds to the
$H > H_\mathrm{vac}$ region of Fig.~\ref{fig:bvp:series:velocity}(b).

The third solution evolves from the most simple initial state: $q = 0$
for each $y$.  In the considered area of small~$f$, such domain walls
prove to be \emph{dynamically} unstable, breaking up in a way similar
to the snake (transverse) instability of the $2 \pi$ domain walls in
atomic BECs that is accompanied by formation of HQV
molecules~\cite{Ihara2019,Gallemi2019}.  However, in contrast to
atomic BECs, the domain wall as such does not disappear but only
reduces its symmetry by turning into a chain of the $q = \pm 1$ states
connected by HQV molecules like in Fig.~\ref{fig:2d:shape}.  The
further evolution is illustrated in Fig.~\ref{fig:2d:chain}.  With
time, adjacent pairs of molecules annihilate or couple into composite
solitons that detach from the wall and decay shortly, producing sound
waves; as a result, the total energy of the system decreases.

\subsection{Spontaneous phenomena}
\label{sec:2d:spontaneous}

\begin{figure}
  \centering
  \includegraphics[width=\linewidth]{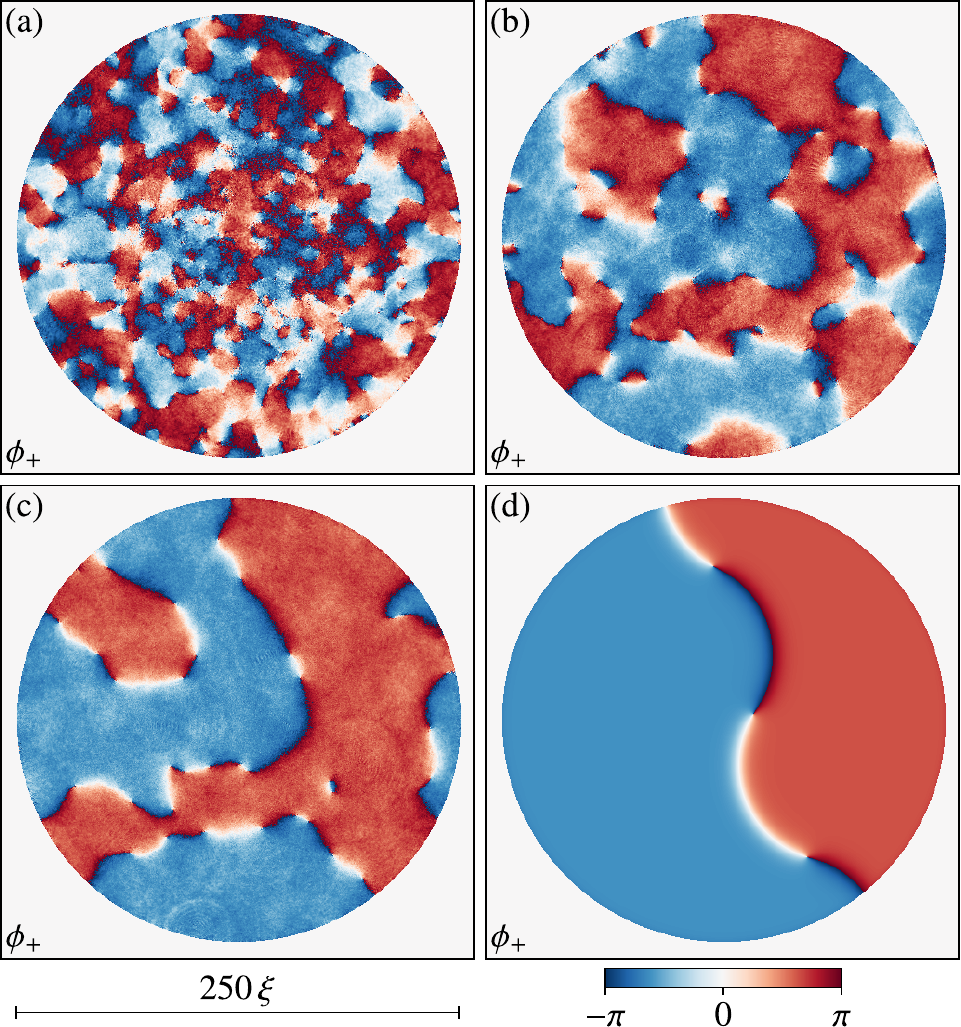}
  \caption{\label{fig:2d:evolution} Typical stages of evolution of a
    system with zero initial conditions under a slow increase of~$f$.
    (a)~Highly disordered state soon after the spin symmetry
    breakdown.  (b)~Onset of phase domains.  (c)~Self-organization of
    nonstationary domains.  (d)~Eventual static solution at
    $f^2 / f_2^2 \approx 0.14$.  The color map represents~$\phi_+$.
    Parameters: $D = \Omega$, $\gamma / \Omega = 5 \times 10^{-3}$,
    $R = 125 \, \xi$.  The dynamics is shown in Supplemental
    Material~V~\cite{Note5}.}
\end{figure}

Up to now, we have focused on the solutions evolved from some special
initial conditions obeying Eqs.~\eqref{eq:soliton}--\eqref{eq:bc2}.
Bearing in mind optically driven microcavities, let us now consider a
system with \emph{zero} initial conditions that evolves owing to an
arbitrarily slow increase in~$f$.  Initially---at very small $f$---all
solutions are spin-symmetric ($\psi_+ = \psi_-$), so that the system
is effectively one-component ($\psi_H \neq 0$, $\psi_V = 0$).
However, as soon as $f$ reaches its critical magnitude
$f_\mathrm{thr}$ \eqref{eq:threshold}, the driven mode is amplified
sharply, which is accompanied by a rapid population of many new modes
with nonzero wave vectors and ``antisymmetric'' spin states such that
$\psi_+ = -\psi_-$ (accordingly, $\psi_H = 0$, $\psi_V \neq 0$).  As a
result, the spin symmetry breaks down and the system comes to a highly
disordered state represented in Fig.~\ref{fig:2d:evolution}(a).

Let us consider this process in more detail.  The initial modulational
instability, occurring at $f \gtrsim f_\mathrm{thr}$ and
$V \psi_H^2 \gtrsim 2 \gamma$, involves a jump in $\psi_H$ up to
$V \psi_H^2 \approx 2 D - \Omega$ that corresponds to the upper root
of Eq.~\eqref{eq:vac:symm}.  Such transition is, in essence, a sharp
switch of a bistable system~\cite{Gavrilov2014.prb.b}.  As a
consequence, the spectra of elementary excitations~\eqref{eq:exc:symm}
take the form
\begin{align}
  \label{eq:exc:H}
  \tilde E_H(k)
  &= -i \gamma \pm \sqrt{E(k) \left[ E(k) + \Omega \right]},
  \\
  \label{eq:exc:V}
  \tilde E_V(k)
  &= -i \gamma \pm \sqrt{E(k) \left[ E(k) - \Omega \right]},
\end{align}
where we assume that $D = \Omega$ for simplicity,
$E(k) = \hbar^2 k^2 / 2 m$ represents the bare dispersion law, and
$H,V$ indicate the polarization states.  $\tilde E_H(k)$ is seen to
have the Bogoliubov form, which means that $\psi_H$ is indeed
increased up to the point where the respective instability is over
$(\Im \tilde E_H = -\gamma)$.  However, the same state shows a strong
instability in~$\psi_V$, owing to which the spin symmetry eventually
breaks down.  The characteristic time of this process $\hbar / \Gamma$
is determined by
$\Gamma = \max \Im \tilde E_V(k) = \Omega/2 - \gamma$, thus, it is
much shorter than the relaxation time $\hbar / \gamma$.  The maximum
of $\Im \tilde E_V(k)$ is reached at $k = \sqrt{m \Omega} / \hbar$,
which results in a significant spatial inhomogeneity.  The respective
scale $k^{-1}$ is as small as $\sqrt{3} \xi$ for $D = \Omega$.  Notice
also that the individual phases of $\psi_V(\pm \bk)$ are not fixed in
accordance with the Bogoliubov equation~\eqref{eq:bdg}, which implies
the spontaneous character of the symmetry breaking.

Since the spin symmetry breaks down rapidly, the overall transition
proceeds \emph{nonglobally:} the signs of~$S_3$ become unrelated
already on the scale of several~$\xi$.  In this respect, the discussed
process agrees with the Kibble-Zurek scenario for the nonadiabatic
symmetry breaking in the early Universe~\cite{Kibble1980} or
superfluid liquids~\cite{Zurek1985}.  The analogy is substantiated by
a direct connection between the characteristic time $2 \hbar / \Omega$
and length scale $\hbar / \sqrt{m \Omega}$ of the transition.
However, the spin symmetry breaking does not necessarily imply
formation of phase domains, because $f_\mathrm{thr} < f_1$ at small
$\gamma$ [see \eqref{eq:threshold}, \eqref{eq:asymm:f1}] and, thus,
some interval of~$f$ has no stable vacuum states at
all~\cite{Gavrilov2022}.

As soon as $f$ reaches $f_1$, a pair of new vacuum states with
$\psi_+^* = \psi_-^\vphast$ come into being.  Since then, the
evolution takes the form of self-organization.  Large-scale domains
start forming at its early stage [Fig.~\ref{fig:2d:evolution}(b)], but
the whole picture also contains a number of moving non-consolidated
fragments similar to Fig.~\ref{fig:2d:drop}.  In fact, even after the
system has divided into domains [Fig.~\ref{fig:2d:evolution}(c)], it
still demonstrates motion of domain walls.  A finite-sized system can
occasionally stabilize completely [Fig.~\ref{fig:2d:evolution}(d)].
According to our calculations, this occurs only when $f$ becomes much
larger than $f_1$ and the $q = \pm 1$ domain walls acquire a
sufficiently high energy.

The system evolution outlined in Fig.~\ref{fig:2d:evolution} is
explicitly shown in Supplemental Material~V~\footnote{See \href
  {https://arxiv.org/src/2505.09553v4/anc/Supplemental_Material_V.mp4}
  {Supplemental Material~V} for a video presentation corresponding to
  Fig.~\ref{fig:2d:evolution}.}.  In this calculation, the pump
intensity $f^2(t)$ was increased linearly during the time period of
$T \approx 450 \,\hbar / \mu$ and then kept constant.  To reduce the
time of the spin symmetry breaking, we added to $f$ at $t < T$ a weak
complex white noise $\tilde f_\pm(\br, t)$ with a fixed amplitude
$|\tilde f_\pm| = \beta$ and zero mean over both $\br$ and $t$.  Since
the symmetry breaks down owing to the exponential growth of
excitations, the time of this process shows only a logarithmic
increase with decreasing~$\beta$.  Thus, the obtained results are
qualitatively independent of $T$ and~$\beta$ (in particular, a
disordered state as in Fig.~\ref{fig:2d:evolution}(a) is inevitable).

\begin{figure}[!b]
  \centering
  \includegraphics[width=\linewidth]{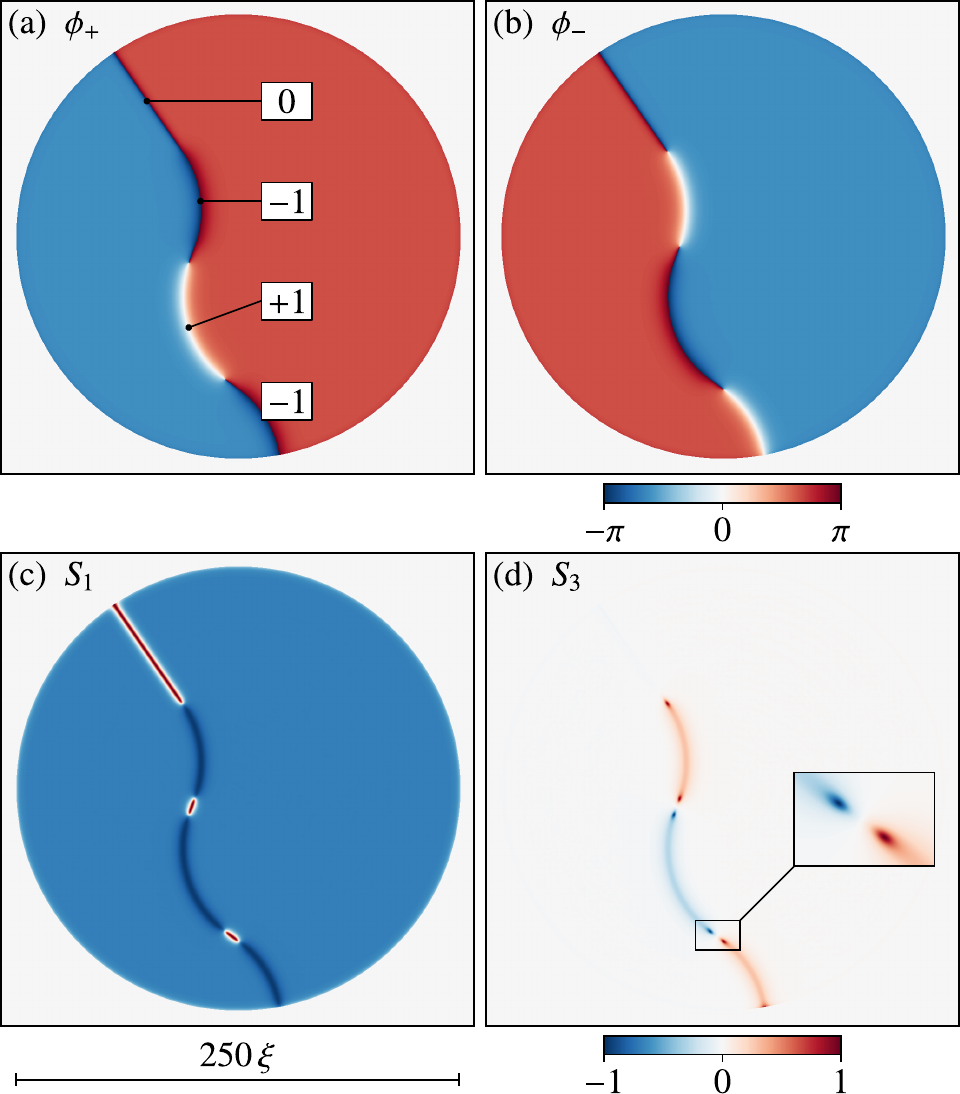}
  \caption{\label{fig:2d:crossover} Solution with coexisting $q = 0$
    and $q = \pm 1$ domain walls.  (a)--(d) represent $\phi_+$,
    $\phi_-$, $S_1$, $S_3$.  Numbers in (a) indicate $q$.  Excitation
    intensity $f^2 / f_2^2 \approx 0.16$, other parameters as in
    Fig.~\ref{fig:2d:evolution}.}
\end{figure}

When $f$ is increased further compared to Fig.~\ref{fig:2d:evolution},
the system exhibits dynamically stable domain walls with $q = 0$.
Notice that even Fig.~\ref{fig:bvp:series:pump} suggests that the
$q = 0$ states become less steep and, thus, less energetically costly
than the $q = \pm 1$ ones as $f$ increases.
Figure~\ref{fig:2d:crossover} shows a crossover solution in which the
two types of domain walls coexist.  It evolved from zero initial
conditions, similarly to Fig.~\ref{fig:2d:evolution}, under
increasing~$f^2$ up to ${\sim} \, 0.16 \, f_2^2$.  The $q = 0$ part of
the domain wall is seen as a straight line segment characterized by
opposite $\phi_+$ and $\phi_-$ [Figs.~\ref{fig:2d:crossover}(a) and
\ref{fig:2d:crossover}(b)], high $S_1$
[Fig.~\ref{fig:2d:crossover}(c)], and almost zero $S_3$
[Fig.~\ref{fig:2d:crossover}(d)].  The rectilinear shape is indicative
of the domain-wall tension that has an energetic origin in analogy to
the $2 \pi$ domain walls in atomic BECs~\cite{Son2002,Eto2018}.

At smaller $f$, all $q = 0$ domain walls were confined inside HQV
molecules much as in Fig.~\ref{fig:2d:shape}.  The same kind of
confined $q = 0$ states is represented in our last example by two tiny
segments with $S_1 \approx +1$ whose end-point HQVs are visible in
Fig.~\ref{fig:2d:crossover}(d).  In turn, the essentially large-scale
$q = 0$ domain walls originate through the \emph{deconfinement} of HQV
pairs that is accompanied by a reordering and possible decay of HQVs
upon increasing~$f$.  This process is, in a way, the opposite to the
snake instability considered previously~[Fig.~\ref{fig:2d:chain}(a)].

Experimentally, phase domains and both types of domain walls can be
distinguished by measuring the Stokes vector components of the emitted
radiation.  The domains \emph{per se} are distinguished by the sign of
$S_2 \approx \pm \sin 2 \alpha$ [Eq.~(9)]; the $q = 0$ domain walls
have $S_1 = +1$ and $S_3 = 0$; the $q = -1$ ($q = +1$) domain walls
have $S_1 \approx -1$ and positive (negative) $S_3$.  These signatures
are clearly seen in Figs.~\ref{fig:2d:crossover}(c) and (d); the signs
of $S_3$ for the $q = \pm 1$ domain walls are in line with
Fig.~\ref{fig:bvp:types}.

\section{Concluding remarks}
\label{sec:conclusion}

We have studied two kinds of phase domain walls formed in a coherently
driven Bose system.  One of them $(q = 0)$ is found to be surprisingly
analogous to domain walls typical of freely evolving BECs, whereas the
other $(q = \pm 1)$ is unconventional.  The spin and spatiotemporal
symmetries of the system break down spontaneously; a pair of its true
vacuum states have opposite phases, neither of which is favored.
Owing to the repulsive interaction and finite lifetime of
quasiparticles, the system tends to reach either of these twin states
at each point.  Since, however, the rate of the symmetry-breaking
transition is great even when~$f$ is increased slowly, its outcome is
spatially dependent, similarly to the Kibble-Zurek scenario.  The
breakup of the trivial vacuum state is somewhat analogous to a
spontaneously emergent polarization in \emph{immiscible} two-component
BECs~\cite{Son2002, Recati2022, Zibold2010, Cominotti2023}, although
it does not rely upon the strong inter-component repulsion typical of
such systems.  Notice also that the spontaneous formation of domain
walls was evidenced earlier in a three-component BEC in a magnetic
field under the situation of a rapid phase transition~\cite{Kang2019}.

Curiously, the special point $f = f_1$ in which the opposite-phase
vacuum states come into being represents a sort of crossover between
the driven and freely evolving BECs.  The spectrum of elementary
excitations at this point takes the Bogoliubov form and the $q = 0$
domain walls show a characteristic $2 \pi$ jump of the relative phase.
In general, the $q = 0$ domain walls acquire nonzero spin polarization
depending on their state of motion (in a one-dimensional system) or
give birth to HQV molecules through the transverse instability in two
dimensions.  Both of these phenomena also occur in atomic
BECs~\cite{Ihara2019,Gallemi2019}.

An important peculiarity of the system considered lies in the fact
that its phase domains \textit{per se} are asymptotically stable.
Dynamical instability of the $q = 0$ domain walls does not end up in a
globally uniform state as in \cite{Ihara2019,Gallemi2019} but produces
new domain walls with broken parity $(q = \pm 1)$.  Such solutions
form a topological doublet whose components differ in the sign of the
total phase variation across the wall.  They tend to move in certain
preferred directions and have fixed polarizations
($\mathop{\mathrm{sgn}} S_3 = -q$), being in this respect similar to
magnetic monopoles.  In two dimensions, the $q = \pm 1$ domain walls
naturally combine into collective states with long-range ordering,
such as spontaneously rotating binary patterns or composite solitons
containing multiple pairs of HQVs.

The obtained results hopefully allow to take a new view of a relation
between atomic BECs and nonlinear optical systems.  Our theoretical
predictions can be directly verified based on the exciton-polariton
fluids in state-of-the-art microcavities.

\acknowledgments

I am grateful to N.~N.~Ipatov for stimulating discussions.  The work
was supported by the Foundation for the Advancement of Theoretical
Physics and Mathematics ``BASIS''.

\appendix

\section{Parameters of the polariton system}
\label{sec:app:parameters}

Optical field in a planar microcavity~\cite{Kavokin.book.2017} has the
shape of a standing wave whose transverse wave number $\pm k_z$ is
fixed but the in-plane wave vector $\bk = (k_x, k_y)$ remains free;
$|\bk|$ corresponds to the angle $\theta$ of incidence or emission of
radiation relative to the $z$ axis: $|\bk / k_z| = \tan \theta$.  In
the range of small $|\bk|$, the dispersion law of cavity photons takes
a parabolic form
\begin{equation}
  \label{eq:disp:photon}
  E_C(\bk) = \frac{\hbar c}{\sqrt{\epsilon}} \sqrt{k_z^2 + \bk^2}
  \approx E_0 + \frac{\hbar^2 \bk^2}{2m_C},
\end{equation}
where $\epsilon$ is dielectric permittivity, $E_0$ is equal to
$\hbar c |k_z| / \sqrt{\epsilon}$, and $m_C = \epsilon E_0 / c^2$ is
the photon mass.  Energy $E_0$ can be tuned to resonance with
quasi-two-dimensional excitons that form in a thin quantum well placed
at the antinode of the electric field.  Since the exciton mass greatly
exceeds $m_C$, the exciton energy $E_X$ is fairly independent of $\bk$
near $\bk = 0$.  The strong exciton-photon coupling, characterized by
the Rabi energy $\mathcal{R}$, involves formation of polaritons.  The
polariton dispersion law
\begin{equation}
  \label{eq:disp:polariton}
  E_\mathrm{LP,UP}(\bk) =
  \frac{E_C(\bk) + E_X}{2} \mp \frac12
  \sqrt{(E_C(\bk) - E_X)^2 + \mathcal{R}^2}
\end{equation}
consists of the lower (LP) and upper (UP) polariton branches.

The external field is assumed to be a plane wave which has zero~$\bk$
(so that its full wave vector is orthogonal to the cavity) and a
frequency that slightly exceeds the respective resonance
$E_\mathrm{LP}(\bk \, {=} \, 0)$.  Since the excitation detuning $D$
is noticeably less than $\mathcal{R}$, the upper polariton branch
never gets populated, so we neglect it.  Then the mean-field
Eqs.~\eqref{eq:mean-field} taking into account the correct form of
$E_\mathrm{LP}(\bk)$ are as follows:
\begin{equation}
  \label{eq:GP}
  i \hbar \frac{\partial \psi_\pm}{\partial t} =
  \left[
    -D + \hat E (-i \nabla) - i \gamma
    + V \psi_\pm^* \psi_\pm^\vphast
  \right] \psi_\pm^\vphast
  + \frac{\Omega}{2} \psi_\mp^\vphast
  + f,
\end{equation}
where operator function $\hat E = \hat E (-i \nabla)$ reproduces the
dispersion law without the ground-state energy,
\begin{equation}
  \label{eq:disp:no-ground}
  E(\bk) = E_\mathrm{LP}(\bk) - E_\mathrm{LP}(0).
\end{equation}
Throughout this work, we employ Eqs.~\eqref{eq:GP} for accurate
dynamical modeling, whereas basic features of domain walls are studied
within the effective-mass approximation.  In particular, we take
$E_X = E_0$ when calculating $E_\mathrm{LP}(\bk)$ in the former case
and, self-consistently, $m = 2 m_C$ in the second case when
Eqs.~\eqref{eq:energy}, \eqref{eq:continuity}, or \eqref{eq:soliton}
are applied.  The deviation between these models is not yet great for
the chosen ratio $D / \mathcal{R} = 0.1$.

The set of parameters used in our numerical examples corresponds to a
GaAs-based microcavity: $\epsilon = 12$, $E_0 = 1.5$\,eV,
$\mathcal{R} = 10$\,meV, $D = \Omega = 1$\,meV; $\gamma$ is small and
chosen arbitrarily.  The only unconventional assumption is that
$\Omega$ largely exceeds~$\gamma$.  Calculations show that phase
domains spontaneously form in some range of $f$ at
$\Omega / \gamma \gtrsim 10$ (see also a discussion
in~\cite{Gavrilov2022}).  However, in order to robustly observe both
types of domain walls, one has to further increase $\Omega / \gamma$
up to ${\sim} \, 10^2$; for instance, Figs.~\ref{fig:2d:shape},
\ref{fig:2d:evolution}, and \ref{fig:2d:crossover} are obtained at
$\Omega / \gamma = 200$.  In practice, high $\Omega / \gamma$ could
originate from a structural anisotropy (e.\,g., owing to an in-plane
mechanical stress \cite{Gavrilov2014.prb.j} or magnetic field)
combined with a high quality factor $Q \approx E_0 / 2 \gamma$.  These
conditions seem to be achievable in the immediate future.

In modeling, any choice of $V > 0$ only fixes the units for $f$ and
$\psi_\pm$ and does not affect the results.  Physically, however, it
is important that the mean-field theory holds true when $V n_\pm$ is
as great as $D$ and $\Omega$.  In accordance with Eq.~(6)
from~\cite{Vladimirova2010},
$V \approx \pi a_B^{\vphantom 2} C_X^2 e^2 / \epsilon \approx 1.5
\times 10^{-14}~\mathrm{eV \, cm}^2$, where $a_B \approx 8$~nm is the
exciton Bohr radius and $C_X^2 = 1/2$ is the $k = 0$ exciton Hopfield
factor for $E_X = E_0$.  Consequently, in the state with
$V n_\pm = \mu = D + \Omega/2 = 1.5$~meV we have
$n_\pm a_B^2 \approx 0.06$, so that the exciton gas is still dilute
and the mean-field approximation is valid~(\cite{Keldysh2017.en}).

The interaction between same-spin excitons or polaritons is, in
essence, Coulombic, so it is much stronger than the $s$-wave
scattering of atoms in conventional BECs.  On the other hand, the
effective mass $m \sim 2 \epsilon E_0 / c^2$ is very small as compared
even to free electrons.  Accordingly, the healing length
$\xi = \hbar / \sqrt{2 m V n_\pm} = \hbar c / \sqrt{4 \epsilon E_0
  \mu} \approx 0.6\,\mu$m turns out to be close to the atomic one, but
the respective sound velocity
$v_s = \sqrt{\mu / m} = c \sqrt{\mu / 2 \epsilon E_0} \approx
0.0065\,c$ is much greater.  As a characteristic example, the
disk-shaped pattern in Fig.~\ref{fig:2d:disk} has the diameter of
0.3~mm and rotates with the period of 0.7~ns.

\section{All vacuum states and elementary excitations}
\label{sec:app:vacuum}

As a first step, one should find amplitudes $\bar \psi_\pm$ of
stationary uniform solutions with $\bk = 0$ (vacuum states).  Provided
that $\partial_t \psi_\pm = 0$ and $\nabla \psi_\pm = 0$,
Eqs.~\eqref{eq:GP} take the form of two coupled algebraic equations
\begin{equation}
  \label{eq:vac}
  \left(
    -D - i \gamma
    + V \bar \psi_\pm^* \bar \psi_\pm^\vphast
  \right) \bar \psi_\pm^\vphast
  + \frac{\Omega}{2} \bar \psi_\mp^\vphast
  + f
  = 0.
\end{equation}
When a particular solution to \eqref{eq:vac} for a given $f$ is found,
the respective spectrum of excitations can be calculated in a standard
way based on the Bogoliubov equation.  Namely, substitution of the
ansatz
\begin{equation}
  \label{eq:bdg:ansatz}
  \psi_\pm(\br, t) = \bar \psi_\pm
  + \zeta_\pm e^{ i ( \bk \br - \tilde E t / \hbar )}
  + \eta_\pm  e^{-i ( \bk \br - \tilde E t / \hbar )}
\end{equation}
with infinitesimal $|\zeta_\pm|$ and $|\eta_\pm|$ into \eqref{eq:GP}
leads one to a linear system
\begin{equation}
  \label{eq:bdg}
  \left[ \,
    \begin{pmatrix}
      A_+
      & B_+
      & \Omega / 2
      & 0
     \\ -B_+^*
      & -A_+^\vphast
      & 0
      & -\Omega / 2
     \\ \Omega / 2
      & 0
      & A_-
      & B_-
     \\ 0
      & -\Omega / 2
      & -B_-^*
      & -A_-^\vphast
    \end{pmatrix}
    - \left( i \gamma + \tilde E \right) I_4
  \right] \,
  \begin{pmatrix}
    \zeta_+ \\ \eta_+^* \\
    \zeta_- \\ \eta_-^*
  \end{pmatrix}
  = 0,
\end{equation}
where
\begin{equation}
  \label{eq:bdg:defs}
  A_\pm = E(\bk) - D + 2 V |\bar \psi_\pm|^2,
  \quad B_\pm = V \bar \psi_\pm^2,
\end{equation}
and $I_4$ is the $4 \times 4$ identity matrix.  Solving the equation
$\det [\cdot] = 0$ yields the sought-for dispersion law
$\tilde E_i = \tilde E_i(\bk)$ that has four branches $(i = 1,2,3,4)$.
Stability of the underlying vacuum states is determined by quantity
\begin{equation}
  \label{eq:increment}
  \Gamma = \max_{\bk, i} \Im \tilde E_i(\bk).
\end{equation}
If $\Gamma < 0$, the solution $\bar \psi_\pm$ to Eqs.~\eqref{eq:vac}
is \emph{asymptotically} stable (attractive), because any small
deviation from it decays with time.  And vice versa, all solutions
with $\Gamma > 0$ are destroyed by arbitrarily small fluctuations and
should thus be considered as false vacuum states.

In general, Eqs.~\eqref{eq:vac} have up to nine solutions for
each~$f$, six of which can be stable.  They fall into three classes.
The first is one or two spin-symmetric solutions such that
$\bar \psi_+ = \bar \psi_-$; the second is a doublet of states with
opposite spin polarizations, $|\bar \psi_+| \neq |\bar \psi_-|$; the
third is a doublet with $\bar \psi_\pm^* = \bar \psi_\mp^\vphast$ at
$\gamma \to 0$ that consists of two solutions with opposite
\emph{phases}.  The last pair turns out to be the sole kind of stable
vacuum states in a finite range of $f$.  In order to prove this
statement conclusively, we should consider all classes of solutions
and analyze their stability.

\textbf{1.} The \emph{spin-symmetric states}
$(\bar \psi_+ = \bar \psi_-)$ exist for each $f$ and are the only
solutions at $f \to 0$ as well as $f \to \infty$.  Taking the absolute
square of~\eqref{eq:vac} readily leads one to
\begin{equation}
  \label{eq:vac:symm}
  f^2 = \frac{u}{2 V}
  \biggl[
    \biggl( D - \frac{\Omega}{2} - \frac{u}{2} \biggr)^2
    + \gamma^2
  \biggr],
\end{equation}
where $u = Vn = 2 V |\bar \psi_\pm|^2$ corresponds to the density of
the full number of particles.  The spectrum of excitations is
\begin{equation}
  \label{eq:exc:symm}
  \tilde E(\bk) = -i \gamma \pm \frac{1}{2}
  \sqrt{\left[ 2 (D - E(\bk) - u) \pm \Omega \right]^2 - u^2}.
\end{equation}
At $u \to 0$ and $f \to 0$, we have $\Gamma = -\gamma$.  Upon
increasing $f$ and $u$, quantity $\Gamma$ changes its sign in the
critical point $u = 2 \gamma$ where the system undergoes modulational
instability.  Specifically, $\Im \tilde E(\bk)$ reaches zero on the
circle $|\bk| = k_\star$ such that
$E(k_\star) = D - \Omega / 2 - 2 \gamma$.  (By definition,
$E(k) \ge 0$, so we suppose that $D - \Omega / 2 > 2 \gamma$.)  The
respective threshold value of $f^2$ is
\begin{equation}
  \label{eq:threshold}
  f_\mathrm{thr}^2 = \frac{\gamma}{V}
  \biggl[
    \biggl( D - \frac{\Omega}{2} - \gamma \biggr)^2
    + \gamma^2
  \biggr].
\end{equation}
Once this threshold is reached, $u$ starts growing spontaneously (even
if $f$ is kept constant) up to the magnitude $u \gtrsim 2 D - \Omega$
that corresponds to a different root of \eqref{eq:vac:symm} for the
same~$f$.  Such a switch between two vacuum states, mediated by
\emph{transient} excitation of many modes with $\bk \neq 0$, was
studied earlier in the case of a scalar bistable
system~\cite{Gavrilov2014.prb.b}.  The two-component system with
$\gamma \ll D \sim \Omega$ behaves differently in that its upper
stationary solution obeying \eqref{eq:vac:symm} is unstable with
respect to the spin symmetry breaking.  Technically, one can make sure
that the eigenstates $\zeta_\pm$ and $\eta_\pm$ corresponding to
$\Im \tilde E > 0$ are antisymmetric ($\zeta_+ = -\zeta_-$,
$\eta_+ = -\eta_-$), so their spontaneous amplification necessarily
breaks the spin symmetry (see Sec.~\ref{sec:2d:spontaneous}).

The second $\Gamma = 0$ point in which the spin-symmetric states
become stable again is reached at
$Vf^2 \approx \Omega^2 (D + \Omega / 2)$.  Notice that the ratio
between this value and $Vf_\mathrm{thr}^2$ occurs to be indefinitely
great at $\gamma \to 0$.

\textbf{2.} The \emph{asymmetric} solutions with
$\bar \psi_+ \neq \bar \psi_-$ exist in a finite area of $u$ and $f$.
Taking the difference of Eqs.~\eqref{eq:vac}, separating the terms
proportional to $\bar \psi_\pm$, and proceeding to their absolute
squares allow one to see that
\begin{equation}
  \label{eq:asymm:product}
  u_+ u_- = (\mu - u)^2 + \gamma^2
  \quad \text{if} \quad
  u_+ \neq u_-,
\end{equation}
where $u_\pm = V n_\pm = V |\bar \psi_\pm|^2$, $u = V n = u_+ + u_-$,
and $\mu = D + \Omega / 2$.  The inequality $(u_+ - u_-)^2 > 0$
together with \eqref{eq:asymm:product} is equivalent to
$l_1 < u < l_2$, where
\begin{equation}
  \label{eq:asymm:limits}
  l_{1,2} = \frac{2}{3} \Bigl(
    2 \mu \mp \sqrt{\mu^2 - 3 \gamma^2}
  \Bigr).
\end{equation}
Employing \eqref{eq:asymm:product} allows one to derive the solutions
to \eqref{eq:vac} in a parametric form.  Specifically, $f^2$ as well
as $\bar \psi_\pm$ can be found as functions of only one real
variable, $u$, defined on the interval $(l_1, l_2)$.  Function
$f^2(u)$ is single-valued and has the form
\begin{equation}
  \label{eq:vac:asymm:f}
  V f^2 =
  \left( 2 \mu - u \right)
  \left( D - u \right)^2
  + \frac{\gamma^2 \left( 2 D - u \right)^2}{2 \mu - u}.
\end{equation}
For each $u$, there is a doublet of solutions which transform into
each other by exchanging $\bar \psi_+$ and $\bar \psi_-$:
$\bar \psi_{\pm(\mp)} = \sqrt{u_{1,2} / V} e^{i \alpha_{1,2}}$, where
$u_{1,2}$ and $e^{i \alpha_{1,2}}$ are determined unambiguously,
\begin{gather}
  \label{eq:vac:asymm:u12}
  u_{1,2} = \frac{u}{2} \pm
  \frac{1}{2} \sqrt{u^2 - 4 [(\mu - u)^2 + \gamma^2]},
  \\
  \label{eq:vac:asymm:phase-diff}
  \alpha_2 - \alpha_1 = \arccos \,
  \Biggl[
  \frac{1}{\sqrt{(\mu - u)^2 + \gamma^2}}
  \Biggl(
  \mu - u + \frac{2 \gamma^2}{2 \mu - u}
  \Biggr)
  \Biggr],
  \\
  \label{eq:vac:asymm:phase-exp}
  e^{i \alpha_{1,2}} = \frac{\sqrt{V f^2}}
  {(D + i \gamma - u_{1,2}) \sqrt{u_{1,2}} -
    (\Omega / 2) \sqrt{u_{2,1}}
    e^{i (\alpha_{2,1} - \alpha_{1,2})}}.
\end{gather}
Notice that $u_2 < u_1$ and $0 < \alpha_2 - \alpha_1 < \pi$.  The
right-hand side of \eqref{eq:vac:asymm:phase-exp} is expressed through
\eqref{eq:vac:asymm:f}--\eqref{eq:vac:asymm:phase-diff} those, in
turn, depend solely on~$u$.  The radicand in \eqref{eq:vac:asymm:u12}
is positive and the argument of arc cosine in
\eqref{eq:vac:asymm:phase-diff} lies in the interval $(-1, 1)$ as long
as $u \in (l_1, l_2)$.

The spectrum of excitations is also parameterized by $u$ and has the
following form,
\begin{multline}
  \label{eq:exc:asymm}
  \tilde E(\bk) =
  -i \gamma
  \pm \frac{1}{\sqrt{2}}
  \Biggl\{
  2 \left( E(\bk) - D + u \right)^2
  + u^2
  - 6 \left( \mu - u \right)^2
  + \frac{g^2}{2}
  - 6 \gamma^2
  \\ {} \pm \Biggl[
  \left(
    u^2 - 4 \left( \mu - u \right)^2 - 4 \gamma^2
  \right)
  \left( 4 E(\bk) - 4 D + 3 u \right)^2
  \\ {} + 4 g^2 \left( E(\bk) - D + u \right)^2
  - g^2 u^2
  \\ {} + 4 g^2 \left(
    \mu - u + \frac{2 \gamma^2}{2 \mu - u}
  \right)^2
  \Biggr]^{1/2}
  \Biggr\}^{1/2}.
\end{multline}
Expressions \eqref{eq:vac:asymm:f}--\eqref{eq:exc:asymm} constitute an
\emph{exact} representation of all asymmetric vacuum states in a
system with arbitrary $D$, $\Omega$, and $\gamma$.

If $\gamma \ll D \sim \Omega$, quartic equation $df / du = 0$ has four
real-valued roots, of which two, $r_{1,2}$, fall within $(l_1, l_2)$.
Let $r_1 < r_2$, then the first relevant interval $l_1 < u < r_1$ has
a positive slope $(du / df > 0)$ and contains the \emph{spin} doublets
whose stability (sign of $\Gamma$) depends on~$f$.  The next one,
$r_1 < u < r_2$, is wholly unstable ($du / df < 0$ and $\Gamma > 0$).
The last interval $r_2 < u < l_2$ contains the opposite-\emph{phase}
doublets which, by contrast, are stable for each~$f$.  At $u \to l_2$,
both $u_+ - u_-$ and $\alpha_+ - \alpha_-$ tend to zero,
\eqref{eq:vac:asymm:f} coincides with \eqref{eq:vac:symm}, and
\eqref{eq:exc:asymm} coincides with \eqref{eq:exc:symm}; moreover,
exactly at this point the spin-symmetric solutions become stable
again, displaying the transcritical bifurcation.

\textbf{2-a.} The \emph{spin} doublets have a moderate amplitude such
that $D < u \sim \mu$, therefore, the last term in
\eqref{eq:vac:asymm:f} can be neglected in the limit $\gamma \to 0$.
This branch of solutions was extensively studied in
Ref.~\cite{Gavrilov2022} where Eq.~\eqref{eq:vac:asymm:f} was first
derived.  Here, we only notice that the spin doublets are unstable at
small~$f$, in contrast to the spin-symmetric states.  The instability
does not necessarily imply scattering into the $\bk \neq 0$ modes.
For instance, at $D = \Omega$, $\gamma \to 0$, and $\bk = 0$, the
energy of excitations \eqref{eq:exc:asymm} becomes
\begin{multline}
  \label{eq:exc:spin-d}
  \tilde E(0) =  -i \gamma \pm \frac{1}{\sqrt{2}}
  \Bigl\{
  \left( u - \Omega \right)
  \left( -3 u + 11 \Omega \right)
  \\ {} \pm \left| u - \Omega \right|
  \Bigl(
  -27 \, u^2 + 126 \, \Omega u - 131 \, \Omega^2
  \Bigr)^{1/2}
  \Bigr\}^{1/2}.
\end{multline}
The inner radicand is negative, resulting in $\Gamma > 0$, for each
$u$ up to $u \approx 1.56 \, \Omega = 1.04 \, \mu$ and, accordingly,
$f^2(u) \approx 0.3 f^2(l_2)$.  The two modes which get amplified at
this critical point in the course of decreasing $f$ are split in
energy, which corresponds to the Hopf bifurcation.

\textbf{2-b.} The \emph{phase} doublets are of prime importance in our
current study.  Below, we show them to exist and be stable even at
$f \to 0$ if $\gamma \to 0$.  Based on Eqs.~\eqref{eq:vac:asymm:u12}
and \eqref{eq:vac:asymm:phase-diff}, we find their spin polarizations
and phases as explicit functions of the driving field.  We identify
the exceptional points in which the spectrum of excitations takes the
Bogoliubov form typical of superfluid liquids and atomic BECs.
Normally, the spectrum has a gap that is also calculated explicitly.

Unlike Ref.~\cite{Gavrilov2022}, we have to take into account the last
term of~\eqref{eq:vac:asymm:f} that plays a major role in spite of
small~$\gamma$.  Indeed, in accordance with \eqref{eq:asymm:limits},
$l_2$ tends to $2 \mu - \gamma^2 / \mu$ at $\gamma \to 0$, thus, that
last term remains finite at $u \to l_2$, whereas the first term
vanishes.

The top branch of solutions begins at $u = r_2$ that is the upper root
of $df / du = 0$ lying within $(l_1, l_2)$.  After substituting
$u = 2 \mu - \rho$, this equation takes the form
\begin{equation}
  \label{eq:asymm:f-derivative}
  3 \rho^4
  - 4 \rho^3 \left( D + \Omega \right)
  + \rho^2 \left[ \left( D + \Omega \right)^2 + \gamma^2 \right]
  -\gamma^2 \Omega^2 = 0,
\end{equation}
and we seek the minimum positive $\rho$ satisfying it.  Clearly, at
small $\gamma$ this is $\rho \approx \gamma \Omega / (D + \Omega)$.
The respective value of $f^2(u)$,
\begin{equation}
  \label{eq:asymm:f1}
  f_1^2 = \frac{2 \gamma \Omega}{V} \left( D + \Omega \right),
\end{equation}
represents the critical intensity of the driving field.  The second
limiting point is simply $f^2(l_2)$, which evaluates to
\begin{equation}
  \label{eq:asymm:f2}
  f_2^2 = \frac{\Omega^2}{V} \left( D + \frac{\Omega}{2} \right).
\end{equation}
It is seen that $f_2 \gg f_1$ at $\gamma \to 0$, however, the overall
change in $u$ throughout the top branch is very small
$({\sim} \, \gamma)$.  Strictly speaking, \eqref{eq:vac:asymm:f} is
senseless at $\gamma = 0$ because of the $0/0$ indeterminacy in its
last term, which means that $u$ becomes independent of $f$.  However,
everything remains regular at $\gamma \to 0^+$.  Below, we always
suppose that $\gamma \ll D \sim \Omega$, thus, we can substitute
$u = 2 \mu - \rho$ directly into \eqref{eq:vac:asymm:f} and then drop
the $\rho^4$ and $\rho^3$ terms, which eventually yields
\begin{equation}
  \label{eq:asymm:u:quadratic}
  u = 2 \mu - \frac{V f^2}{2 \left( D + \Omega \right)^2}
  \, \Biggl( 1 - \sqrt{1 - \frac{f_1^4}{f^4}} \Biggr)
\end{equation}
(the second root would match the unstable branch of solutions with
$du / df < 0$).  In the case when $f^4 \gg f_1^4$, this result can be
simplified further to yield
\begin{equation}
  \label{eq:asymm:u}
  u = 2 \mu - \frac{\gamma^2 \Omega^2}{V f^2}.
\end{equation}
Substituting \eqref{eq:asymm:u} into \eqref{eq:vac:asymm:u12} and
\eqref{eq:vac:asymm:phase-diff} allows one to find the spin
polarization $S_3$ and phases at high $f$ along the top branch.  In
particular,
\begin{equation}
  \label{eq:asymm:s3}
  S_3 = \pm \frac{\gamma}{\mu} \sqrt{\frac{f_2^2}{f^2} - 1}.
\end{equation}
It turns out that $u_+$ and $u_-$ are very close almost irrespective
of $f$ (even the maximum $|S_3|$ reached at $f = f_1$ is of the order
of $\sqrt{\gamma / \mu}$).  On the other hand, it is clear from
\eqref{eq:vac:asymm:phase-exp} that condition $u_+ = u_-$ necessarily
implies $e^{i \alpha_+} = e^{-i \alpha_-}$ as long as $\gamma \to 0$.
As a result, from \eqref{eq:vac:asymm:phase-diff} and
\eqref{eq:vac:asymm:phase-exp} we have
\begin{equation}
  \label{eq:asymm:phase}
  \alpha_{\pm(\mp)} = \pm \arccos \, \left( -\frac{f}{f_2} \right).
\end{equation}
for two solutions in the phase doublet (generally, the sign of
$\sin(\alpha_- - \alpha_+)$ is equal to the sign of $S_3$).  The fact
that both components are $\pi$-shifted relative to the driving field
at $f = f_2$ agrees with the spin-symmetric model at great $f$
and~$u$.

Proceeding to the problem of excitations, notice that everything in
\eqref{eq:exc:asymm} hardly depends on $f$ except the last
(fractional) term that can be rewritten as
\begin{equation}
  \label{asymm:fraction}
  \frac{\gamma^2}{2 \mu - u}
  = \frac{V f^2}{\Omega^2}
  = \frac{\mu f^2}{f_2^2}
\end{equation}
in accordance with \eqref{eq:asymm:u} and \eqref{eq:asymm:f2}.  We
substitute \eqref{asymm:fraction} into \eqref{eq:exc:asymm} and assume
that $u = 2 \mu$ in all regular terms, which leads us to the following
explicit form of the spectrum,
\begin{multline}
  \label{eq:exc:top}
  \tilde E(\bk) = -i \gamma
  \pm \Biggl\{
  \left( E(\bk) + D + \Omega \right)^2
  -D \left(D + \Omega \right)
  \\ {} \pm \Omega \, \Biggl[
  \left( E(\bk) + D + \Omega \right)^2
  -4 \mu^2 \,
  \Biggl( \frac{f^2}{f_2^2} - \frac{f^4}{f_2^4} \Biggr)
  \Biggr]^{1/2}
  \Biggr\}^{1/2}.
\end{multline}
It is seen that $\Gamma = -\gamma$ for each $f$, i.\,e., all phase
doublets are stable, which is intimately connected to the fact that
the spectrum has a gap $E_g = \min_{\bk, i} |\Re \tilde E_i(\bk)|$.
The minimum over $\bk$ is reached at $\bk = 0$, so the gap width
evaluates to
\begin{equation}
  \label{eq:exc:asymm:gap}
  E_g = \sqrt{\Omega} \,
  \Biggl\{
  D + \Omega
  -\Biggl[
  \left( D + \Omega \right)^2
  -4 \mu^2 \,
  \Biggl( \frac{f^2}{f_2^2} - \frac{f^4}{f_2^4} \Biggr)
  \Biggr]^{1/2}
  \Biggr\}^{1/2}.
\end{equation}
Clearly, $E_g$ tends to zero at $f = f_1 \propto \sqrt{\gamma} \to 0$
as well as $f = f_2$.  In both of these points, the gapless branches
of \eqref{eq:exc:top} take the Bogoliubov form
\begin{equation}
  \label{eq:exc:bgl}
  \tilde E(\bk) = -i \gamma
  \pm \sqrt{E(\bk) \left[ E(\bk) + 2 \mu \right]}.
\end{equation}
Thus, $\mu \equiv D + \Omega / 2$ matches the chemical potential of an
equilibrium BEC.


\begin{thebibliography}{61}%
\makeatletter
\providecommand \@ifxundefined [1]{%
 \@ifx{#1\undefined}
}%
\providecommand \@ifnum [1]{%
 \ifnum #1\expandafter \@firstoftwo
 \else \expandafter \@secondoftwo
 \fi
}%
\providecommand \@ifx [1]{%
 \ifx #1\expandafter \@firstoftwo
 \else \expandafter \@secondoftwo
 \fi
}%
\providecommand \natexlab [1]{#1}%
\providecommand \enquote  [1]{``#1''}%
\providecommand \bibnamefont  [1]{#1}%
\providecommand \bibfnamefont [1]{#1}%
\providecommand \citenamefont [1]{#1}%
\providecommand \href@noop [0]{\@secondoftwo}%
\providecommand \href [0]{\begingroup \@sanitize@url \@href}%
\providecommand \@href[1]{\@@startlink{#1}\@@href}%
\providecommand \@@href[1]{\endgroup#1\@@endlink}%
\providecommand \@sanitize@url [0]{\catcode `\\12\catcode `\$12\catcode
  `\&12\catcode `\#12\catcode `\^12\catcode `\_12\catcode `\%12\relax}%
\providecommand \@@startlink[1]{}%
\providecommand \@@endlink[0]{}%
\providecommand \url  [0]{\begingroup\@sanitize@url \@url }%
\providecommand \@url [1]{\endgroup\@href {#1}{\urlprefix }}%
\providecommand \urlprefix  [0]{URL }%
\providecommand \Eprint [0]{\href }%
\providecommand \doibase [0]{https://doi.org/}%
\providecommand \selectlanguage [0]{\@gobble}%
\providecommand \bibinfo  [0]{\@secondoftwo}%
\providecommand \bibfield  [0]{\@secondoftwo}%
\providecommand \translation [1]{[#1]}%
\providecommand \BibitemOpen [0]{}%
\providecommand \bibitemStop [0]{}%
\providecommand \bibitemNoStop [0]{.\EOS\space}%
\providecommand \EOS [0]{\spacefactor3000\relax}%
\providecommand \BibitemShut  [1]{\csname bibitem#1\endcsname}%
\let\auto@bib@innerbib\@empty
\bibitem [{\citenamefont {Leggett}(2001)}]{Leggett2001}%
  \BibitemOpen
  \bibfield  {author} {\bibinfo {author} {\bibfnamefont {A.~J.}\ \bibnamefont
  {Leggett}},\ }\bibfield  {title} {\bibinfo {title} {{Bose-Einstein
  condensation in the alkali gases: Some fundamental concepts}},\ }\href
  {https://doi.org/10.1103/RevModPhys.73.307} {\bibfield  {journal} {\bibinfo
  {journal} {Rev. Mod. Phys.}\ }\textbf {\bibinfo {volume} {73}},\ \bibinfo
  {pages} {307} (\bibinfo {year} {2001})}\BibitemShut {NoStop}%
\bibitem [{\citenamefont {Tanaka}(2001)}]{Tanaka2001}%
  \BibitemOpen
  \bibfield  {author} {\bibinfo {author} {\bibfnamefont {Y.}~\bibnamefont
  {Tanaka}},\ }\bibfield  {title} {\bibinfo {title} {{Soliton in Two-Band
  Superconductor}},\ }\href {https://doi.org/10.1103/PhysRevLett.88.017002}
  {\bibfield  {journal} {\bibinfo  {journal} {Phys. Rev. Lett.}\ }\textbf
  {\bibinfo {volume} {88}},\ \bibinfo {pages} {017002} (\bibinfo {year}
  {2001})}\BibitemShut {NoStop}%
\bibitem [{\citenamefont {Son}\ and\ \citenamefont
  {Stephanov}(2002)}]{Son2002}%
  \BibitemOpen
  \bibfield  {author} {\bibinfo {author} {\bibfnamefont {D.~T.}\ \bibnamefont
  {Son}}\ and\ \bibinfo {author} {\bibfnamefont {M.~A.}\ \bibnamefont
  {Stephanov}},\ }\bibfield  {title} {\bibinfo {title} {{Domain walls of
  relative phase in two-component Bose-Einstein condensates}},\ }\href
  {https://doi.org/10.1103/PhysRevA.65.063621} {\bibfield  {journal} {\bibinfo
  {journal} {Phys. Rev. A}\ }\textbf {\bibinfo {volume} {65}},\ \bibinfo
  {pages} {063621} (\bibinfo {year} {2002})}\BibitemShut {NoStop}%
\bibitem [{\citenamefont {Eto}\ \emph {et~al.}(2011)\citenamefont {Eto},
  \citenamefont {Kasamatsu}, \citenamefont {Nitta}, \citenamefont {Takeuchi},\
  and\ \citenamefont {Tsubota}}]{Eto2011}%
  \BibitemOpen
  \bibfield  {author} {\bibinfo {author} {\bibfnamefont {M.}~\bibnamefont
  {Eto}}, \bibinfo {author} {\bibfnamefont {K.}~\bibnamefont {Kasamatsu}},
  \bibinfo {author} {\bibfnamefont {M.}~\bibnamefont {Nitta}}, \bibinfo
  {author} {\bibfnamefont {H.}~\bibnamefont {Takeuchi}},\ and\ \bibinfo
  {author} {\bibfnamefont {M.}~\bibnamefont {Tsubota}},\ }\bibfield  {title}
  {\bibinfo {title} {{Interaction of half-quantized vortices in two-component
  Bose-Einstein condensates}},\ }\href
  {https://doi.org/10.1103/PhysRevA.83.063603} {\bibfield  {journal} {\bibinfo
  {journal} {Phys. Rev. A}\ }\textbf {\bibinfo {volume} {83}},\ \bibinfo
  {pages} {063603} (\bibinfo {year} {2011})}\BibitemShut {NoStop}%
\bibitem [{\citenamefont {Tylutki}\ \emph {et~al.}(2016)\citenamefont
  {Tylutki}, \citenamefont {Pitaevskii}, \citenamefont {Recati},\ and\
  \citenamefont {Stringari}}]{Tylutki2016}%
  \BibitemOpen
  \bibfield  {author} {\bibinfo {author} {\bibfnamefont {M.}~\bibnamefont
  {Tylutki}}, \bibinfo {author} {\bibfnamefont {L.~P.}\ \bibnamefont
  {Pitaevskii}}, \bibinfo {author} {\bibfnamefont {A.}~\bibnamefont {Recati}},\
  and\ \bibinfo {author} {\bibfnamefont {S.}~\bibnamefont {Stringari}},\
  }\bibfield  {title} {\bibinfo {title} {{Confinement and precession of vortex
  pairs in coherently coupled Bose-Einstein condensates}},\ }\href
  {https://doi.org/10.1103/PhysRevA.93.043623} {\bibfield  {journal} {\bibinfo
  {journal} {Phys. Rev. A}\ }\textbf {\bibinfo {volume} {93}},\ \bibinfo
  {pages} {043623} (\bibinfo {year} {2016})}\BibitemShut {NoStop}%
\bibitem [{\citenamefont {Eto}\ and\ \citenamefont {Nitta}(2018)}]{Eto2018}%
  \BibitemOpen
  \bibfield  {author} {\bibinfo {author} {\bibfnamefont {M.}~\bibnamefont
  {Eto}}\ and\ \bibinfo {author} {\bibfnamefont {M.}~\bibnamefont {Nitta}},\
  }\bibfield  {title} {\bibinfo {title} {{Confinement of half-quantized
  vortices in coherently coupled Bose-Einstein condensates: Simulating quark
  confinement in a QCD-like theory}},\ }\href
  {https://doi.org/10.1103/PhysRevA.97.023613} {\bibfield  {journal} {\bibinfo
  {journal} {Phys. Rev. A}\ }\textbf {\bibinfo {volume} {97}},\ \bibinfo
  {pages} {023613} (\bibinfo {year} {2018})}\BibitemShut {NoStop}%
\bibitem [{\citenamefont {Eto}\ \emph {et~al.}(2020)\citenamefont {Eto},
  \citenamefont {Ikeno},\ and\ \citenamefont {Nitta}}]{Eto2020}%
  \BibitemOpen
  \bibfield  {author} {\bibinfo {author} {\bibfnamefont {M.}~\bibnamefont
  {Eto}}, \bibinfo {author} {\bibfnamefont {K.}~\bibnamefont {Ikeno}},\ and\
  \bibinfo {author} {\bibfnamefont {M.}~\bibnamefont {Nitta}},\ }\bibfield
  {title} {\bibinfo {title} {{Collision dynamics and reactions of fractional
  vortex molecules in coherently coupled Bose-Einstein condensates}},\ }\href
  {https://doi.org/10.1103/PhysRevResearch.2.033373} {\bibfield  {journal}
  {\bibinfo  {journal} {Phys. Rev. Res.}\ }\textbf {\bibinfo {volume} {2}},\
  \bibinfo {pages} {033373} (\bibinfo {year} {2020})}\BibitemShut {NoStop}%
\bibitem [{\citenamefont {Recati}\ and\ \citenamefont
  {Stringari}(2022)}]{Recati2022}%
  \BibitemOpen
  \bibfield  {author} {\bibinfo {author} {\bibfnamefont {A.}~\bibnamefont
  {Recati}}\ and\ \bibinfo {author} {\bibfnamefont {S.}~\bibnamefont
  {Stringari}},\ }\bibfield  {title} {\bibinfo {title} {{Coherently Coupled
  Mixtures of Ultracold Atomic Gases}},\ }\href
  {https://doi.org/10.1146/annurev-conmatphys-031820-121316} {\bibfield
  {journal} {\bibinfo  {journal} {Annu. Rev. Condens. Matter Phys.}\ }\textbf
  {\bibinfo {volume} {13}},\ \bibinfo {pages} {407} (\bibinfo {year}
  {2022})}\BibitemShut {NoStop}%
\bibitem [{\citenamefont {M{\"a}kinen}\ \emph {et~al.}(2019)\citenamefont
  {M{\"a}kinen}, \citenamefont {Dmitriev}, \citenamefont {Nissinen},
  \citenamefont {Rysti}, \citenamefont {Volovik}, \citenamefont {Yudin},
  \citenamefont {Zhang},\ and\ \citenamefont {Eltsov}}]{Makinen2019}%
  \BibitemOpen
  \bibfield  {author} {\bibinfo {author} {\bibfnamefont {J.~T.}\ \bibnamefont
  {M{\"a}kinen}}, \bibinfo {author} {\bibfnamefont {V.~V.}\ \bibnamefont
  {Dmitriev}}, \bibinfo {author} {\bibfnamefont {J.}~\bibnamefont {Nissinen}},
  \bibinfo {author} {\bibfnamefont {J.}~\bibnamefont {Rysti}}, \bibinfo
  {author} {\bibfnamefont {G.~E.}\ \bibnamefont {Volovik}}, \bibinfo {author}
  {\bibfnamefont {A.~N.}\ \bibnamefont {Yudin}}, \bibinfo {author}
  {\bibfnamefont {K.}~\bibnamefont {Zhang}},\ and\ \bibinfo {author}
  {\bibfnamefont {V.~B.}\ \bibnamefont {Eltsov}},\ }\bibfield  {title}
  {\bibinfo {title} {{Half-quantum vortices and walls bounded by strings in the
  polar-distorted phases of topological superfluid ${}^3$He}},\ }\href
  {https://doi.org/10.1038/s41467-018-08204-8} {\bibfield  {journal} {\bibinfo
  {journal} {Nat. Commun.}\ }\textbf {\bibinfo {volume} {10}},\ \bibinfo
  {pages} {237} (\bibinfo {year} {2019})}\BibitemShut {NoStop}%
\bibitem [{\citenamefont {Tanaka}(2015)}]{Tanaka2015}%
  \BibitemOpen
  \bibfield  {author} {\bibinfo {author} {\bibfnamefont {Y.}~\bibnamefont
  {Tanaka}},\ }\bibfield  {title} {\bibinfo {title} {{Multicomponent
  superconductivity based on multiband superconductors}},\ }\href
  {https://doi.org/10.1088/0953-2048/28/3/034002} {\bibfield  {journal}
  {\bibinfo  {journal} {Supercond. Sci. Technol.}\ }\textbf {\bibinfo {volume}
  {28}},\ \bibinfo {pages} {034002} (\bibinfo {year} {2015})}\BibitemShut
  {NoStop}%
\bibitem [{\citenamefont {Kibble}\ \emph {et~al.}(1982)\citenamefont {Kibble},
  \citenamefont {Lazarides},\ and\ \citenamefont {Shafi}}]{Kibble1982}%
  \BibitemOpen
  \bibfield  {author} {\bibinfo {author} {\bibfnamefont {T.~W.~B.}\
  \bibnamefont {Kibble}}, \bibinfo {author} {\bibfnamefont {G.}~\bibnamefont
  {Lazarides}},\ and\ \bibinfo {author} {\bibfnamefont {Q.}~\bibnamefont
  {Shafi}},\ }\bibfield  {title} {\bibinfo {title} {{Walls bounded by
  strings}},\ }\href {https://doi.org/10.1103/PhysRevD.26.435} {\bibfield
  {journal} {\bibinfo  {journal} {Phys. Rev. D}\ }\textbf {\bibinfo {volume}
  {26}},\ \bibinfo {pages} {435} (\bibinfo {year} {1982})}\BibitemShut
  {NoStop}%
\bibitem [{\citenamefont {Rubakov}\ and\ \citenamefont
  {Shaposhnikov}(1983)}]{Rubakov1983}%
  \BibitemOpen
  \bibfield  {author} {\bibinfo {author} {\bibfnamefont {V.~A.}\ \bibnamefont
  {Rubakov}}\ and\ \bibinfo {author} {\bibfnamefont {M.~E.}\ \bibnamefont
  {Shaposhnikov}},\ }\bibfield  {title} {\bibinfo {title} {{Do we live inside a
  domain wall?}},\ }\href {https://doi.org/10.1016/0370-2693(83)91253-4}
  {\bibfield  {journal} {\bibinfo  {journal} {Phys. Lett. B}\ }\textbf
  {\bibinfo {volume} {125}},\ \bibinfo {pages} {136} (\bibinfo {year}
  {1983})}\BibitemShut {NoStop}%
\bibitem [{\citenamefont {Kibble}(1980)}]{Kibble1980}%
  \BibitemOpen
  \bibfield  {author} {\bibinfo {author} {\bibfnamefont {T.~W.~B.}\
  \bibnamefont {Kibble}},\ }\bibfield  {title} {\bibinfo {title} {{Some
  implications of a cosmological phase transition}},\ }\href
  {https://doi.org/10.1016/0370-1573(80)90091-5} {\bibfield  {journal}
  {\bibinfo  {journal} {Phys. Rep.}\ }\textbf {\bibinfo {volume} {67}},\
  \bibinfo {pages} {183} (\bibinfo {year} {1980})}\BibitemShut {NoStop}%
\bibitem [{\citenamefont {Zurek}(1985)}]{Zurek1985}%
  \BibitemOpen
  \bibfield  {author} {\bibinfo {author} {\bibfnamefont {W.~H.}\ \bibnamefont
  {Zurek}},\ }\bibfield  {title} {\bibinfo {title} {{Cosmological experiments
  in superfluid helium?}},\ }\href {https://doi.org/10.1038/317505a0}
  {\bibfield  {journal} {\bibinfo  {journal} {Nature}\ }\textbf {\bibinfo
  {volume} {317}},\ \bibinfo {pages} {505} (\bibinfo {year}
  {1985})}\BibitemShut {NoStop}%
\bibitem [{\citenamefont {Kang}\ \emph {et~al.}(2019)\citenamefont {Kang},
  \citenamefont {Seo}, \citenamefont {Takeuchi},\ and\ \citenamefont
  {Shin}}]{Kang2019}%
  \BibitemOpen
  \bibfield  {author} {\bibinfo {author} {\bibfnamefont {S.}~\bibnamefont
  {Kang}}, \bibinfo {author} {\bibfnamefont {S.~W.}\ \bibnamefont {Seo}},
  \bibinfo {author} {\bibfnamefont {H.}~\bibnamefont {Takeuchi}},\ and\
  \bibinfo {author} {\bibfnamefont {Y.}~\bibnamefont {Shin}},\ }\bibfield
  {title} {\bibinfo {title} {{Observation of Wall-Vortex Composite Defects in a
  Spinor Bose-Einstein Condensate}},\ }\href
  {https://doi.org/10.1103/PhysRevLett.122.095301} {\bibfield  {journal}
  {\bibinfo  {journal} {Phys. Rev. Lett.}\ }\textbf {\bibinfo {volume} {122}},\
  \bibinfo {pages} {095301} (\bibinfo {year} {2019})}\BibitemShut {NoStop}%
\bibitem [{\citenamefont {Kavokin}\ \emph {et~al.}(2017)\citenamefont
  {Kavokin}, \citenamefont {Baumberg}, \citenamefont {Malpuech},\ and\
  \citenamefont {Laussy}}]{Kavokin.book.2017}%
  \BibitemOpen
  \bibfield  {author} {\bibinfo {author} {\bibfnamefont {A.~V.}\ \bibnamefont
  {Kavokin}}, \bibinfo {author} {\bibfnamefont {J.~J.}\ \bibnamefont
  {Baumberg}}, \bibinfo {author} {\bibfnamefont {G.}~\bibnamefont {Malpuech}},\
  and\ \bibinfo {author} {\bibfnamefont {P.}~\bibnamefont {Laussy}},\
  }\href@noop {} {\emph {\bibinfo {title} {{Microcavities}}}},\ \bibinfo
  {edition} {2nd}\ ed.\ (\bibinfo  {publisher} {Oxford University Press},\
  \bibinfo {address} {New York},\ \bibinfo {year} {2017})\BibitemShut {NoStop}%
\bibitem [{\citenamefont {Gavrilov}(2020{\natexlab{a}})}]{Gavrilov2020.usp.en}%
  \BibitemOpen
  \bibfield  {author} {\bibinfo {author} {\bibfnamefont {S.~S.}\ \bibnamefont
  {Gavrilov}},\ }\bibfield  {title} {\bibinfo {title} {{Nonequilibrium
  transitions, chaos, and chimera states in exciton-polariton systems}},\
  }\href {https://doi.org/10.3367/ufne.2019.04.038549} {\bibfield  {journal}
  {\bibinfo  {journal} {Phys. Usp.}\ }\textbf {\bibinfo {volume} {63}},\
  \bibinfo {pages} {123} (\bibinfo {year} {2020}{\natexlab{a}})}\BibitemShut
  {NoStop}%
\bibitem [{\citenamefont {Kasprzak}\ \emph {et~al.}(2006)\citenamefont
  {Kasprzak}, \citenamefont {Richard}, \citenamefont {Kundermann},
  \citenamefont {Baas}, \citenamefont {Jeambrun}, \citenamefont {Keeling},
  \citenamefont {Marchetti}, \citenamefont {Szyma{\'n}ska}, \citenamefont
  {Andr{\'e}}, \citenamefont {Staehli}, \citenamefont {Savona}, \citenamefont
  {Littlewood}, \citenamefont {Deveaud},\ and\ \citenamefont
  {Dang}}]{Kasprzak2006}%
  \BibitemOpen
  \bibfield  {author} {\bibinfo {author} {\bibfnamefont {J.}~\bibnamefont
  {Kasprzak}}, \bibinfo {author} {\bibfnamefont {M.}~\bibnamefont {Richard}},
  \bibinfo {author} {\bibfnamefont {S.}~\bibnamefont {Kundermann}}, \bibinfo
  {author} {\bibfnamefont {A.}~\bibnamefont {Baas}}, \bibinfo {author}
  {\bibfnamefont {P.}~\bibnamefont {Jeambrun}}, \bibinfo {author}
  {\bibfnamefont {J.~M.~J.}\ \bibnamefont {Keeling}}, \bibinfo {author}
  {\bibfnamefont {F.~M.}\ \bibnamefont {Marchetti}}, \bibinfo {author}
  {\bibfnamefont {M.~H.}\ \bibnamefont {Szyma{\'n}ska}}, \bibinfo {author}
  {\bibfnamefont {R.}~\bibnamefont {Andr{\'e}}}, \bibinfo {author}
  {\bibfnamefont {J.~L.}\ \bibnamefont {Staehli}}, \bibinfo {author}
  {\bibfnamefont {V.}~\bibnamefont {Savona}}, \bibinfo {author} {\bibfnamefont
  {P.~B.}\ \bibnamefont {Littlewood}}, \bibinfo {author} {\bibfnamefont
  {B.}~\bibnamefont {Deveaud}},\ and\ \bibinfo {author} {\bibfnamefont {L.~S.}\
  \bibnamefont {Dang}},\ }\bibfield  {title} {\bibinfo {title} {{Bose-Einstein
  condensation of exciton polaritons}},\ }\href
  {https://doi.org/10.1038/nature05131} {\bibfield  {journal} {\bibinfo
  {journal} {Nature}\ }\textbf {\bibinfo {volume} {443}},\ \bibinfo {pages}
  {409} (\bibinfo {year} {2006})}\BibitemShut {NoStop}%
\bibitem [{\citenamefont {Baas}\ \emph {et~al.}(2006)\citenamefont {Baas},
  \citenamefont {Karr}, \citenamefont {Romanelli}, \citenamefont {Bramati},\
  and\ \citenamefont {Giacobino}}]{Baas2006}%
  \BibitemOpen
  \bibfield  {author} {\bibinfo {author} {\bibfnamefont {A.}~\bibnamefont
  {Baas}}, \bibinfo {author} {\bibfnamefont {J.-P.}\ \bibnamefont {Karr}},
  \bibinfo {author} {\bibfnamefont {M.}~\bibnamefont {Romanelli}}, \bibinfo
  {author} {\bibfnamefont {A.}~\bibnamefont {Bramati}},\ and\ \bibinfo {author}
  {\bibfnamefont {E.}~\bibnamefont {Giacobino}},\ }\bibfield  {title} {\bibinfo
  {title} {{Quantum Degeneracy of Microcavity Polaritons}},\ }\href
  {https://doi.org/10.1103/PhysRevLett.96.176401} {\bibfield  {journal}
  {\bibinfo  {journal} {Phys. Rev. Lett.}\ }\textbf {\bibinfo {volume} {96}},\
  \bibinfo {pages} {176401} (\bibinfo {year} {2006})}\BibitemShut {NoStop}%
\bibitem [{\citenamefont {Ciuti}\ \emph {et~al.}(1998)\citenamefont {Ciuti},
  \citenamefont {Savona}, \citenamefont {Piermarocchi}, \citenamefont
  {Quattropani},\ and\ \citenamefont {Schwendimann}}]{Ciuti1998}%
  \BibitemOpen
  \bibfield  {author} {\bibinfo {author} {\bibfnamefont {C.}~\bibnamefont
  {Ciuti}}, \bibinfo {author} {\bibfnamefont {V.}~\bibnamefont {Savona}},
  \bibinfo {author} {\bibfnamefont {C.}~\bibnamefont {Piermarocchi}}, \bibinfo
  {author} {\bibfnamefont {A.}~\bibnamefont {Quattropani}},\ and\ \bibinfo
  {author} {\bibfnamefont {P.}~\bibnamefont {Schwendimann}},\ }\bibfield
  {title} {\bibinfo {title} {{Role of the exchange of carriers in elastic
  exciton-exciton scattering in quantum wells}},\ }\href
  {https://doi.org/10.1103/PhysRevB.58.7926} {\bibfield  {journal} {\bibinfo
  {journal} {Phys. Rev. B}\ }\textbf {\bibinfo {volume} {58}},\ \bibinfo
  {pages} {7926} (\bibinfo {year} {1998})}\BibitemShut {NoStop}%
\bibitem [{\citenamefont {Vladimirova}\ \emph {et~al.}(2010)\citenamefont
  {Vladimirova}, \citenamefont {Cronenberger}, \citenamefont {Scalbert},
  \citenamefont {Kavokin}, \citenamefont {Miard}, \citenamefont
  {Lema{\^\i}{}tre}, \citenamefont {Bloch}, \citenamefont {Solnyshkov},
  \citenamefont {Malpuech},\ and\ \citenamefont {Kavokin}}]{Vladimirova2010}%
  \BibitemOpen
  \bibfield  {author} {\bibinfo {author} {\bibfnamefont {M.}~\bibnamefont
  {Vladimirova}}, \bibinfo {author} {\bibfnamefont {S.}~\bibnamefont
  {Cronenberger}}, \bibinfo {author} {\bibfnamefont {D.}~\bibnamefont
  {Scalbert}}, \bibinfo {author} {\bibfnamefont {K.~V.}\ \bibnamefont
  {Kavokin}}, \bibinfo {author} {\bibfnamefont {A.}~\bibnamefont {Miard}},
  \bibinfo {author} {\bibfnamefont {A.}~\bibnamefont {Lema{\^\i}{}tre}},
  \bibinfo {author} {\bibfnamefont {J.}~\bibnamefont {Bloch}}, \bibinfo
  {author} {\bibfnamefont {D.}~\bibnamefont {Solnyshkov}}, \bibinfo {author}
  {\bibfnamefont {G.}~\bibnamefont {Malpuech}},\ and\ \bibinfo {author}
  {\bibfnamefont {A.~V.}\ \bibnamefont {Kavokin}},\ }\bibfield  {title}
  {\bibinfo {title} {{Polariton-polariton interaction constants in
  microcavities}},\ }\href {https://doi.org/10.1103/PhysRevB.82.075301}
  {\bibfield  {journal} {\bibinfo  {journal} {Phys. Rev. B}\ }\textbf {\bibinfo
  {volume} {82}},\ \bibinfo {pages} {075301} (\bibinfo {year}
  {2010})}\BibitemShut {NoStop}%
\bibitem [{\citenamefont {Sekretenko}\ \emph {et~al.}(2013)\citenamefont
  {Sekretenko}, \citenamefont {Gavrilov},\ and\ \citenamefont
  {Kulakovskii}}]{Sekretenko2013.10ps}%
  \BibitemOpen
  \bibfield  {author} {\bibinfo {author} {\bibfnamefont {A.~V.}\ \bibnamefont
  {Sekretenko}}, \bibinfo {author} {\bibfnamefont {S.~S.}\ \bibnamefont
  {Gavrilov}},\ and\ \bibinfo {author} {\bibfnamefont {V.~D.}\ \bibnamefont
  {Kulakovskii}},\ }\bibfield  {title} {\bibinfo {title} {{Polariton-polariton
  interactions in microcavities under a resonant 10 to 100 picosecond pulse
  excitation}},\ }\href {https://doi.org/10.1103/PhysRevB.88.195302} {\bibfield
   {journal} {\bibinfo  {journal} {Phys. Rev. B}\ }\textbf {\bibinfo {volume}
  {88}},\ \bibinfo {pages} {195302} (\bibinfo {year} {2013})}\BibitemShut
  {NoStop}%
\bibitem [{\citenamefont {Gavrilov}\ \emph {et~al.}(2014)\citenamefont
  {Gavrilov}, \citenamefont {Brichkin}, \citenamefont {Novikov}, \citenamefont
  {H{\"o}fling}, \citenamefont {Schneider}, \citenamefont {Kamp}, \citenamefont
  {Forchel},\ and\ \citenamefont {Kulakovskii}}]{Gavrilov2014.prb.j}%
  \BibitemOpen
  \bibfield  {author} {\bibinfo {author} {\bibfnamefont {S.~S.}\ \bibnamefont
  {Gavrilov}}, \bibinfo {author} {\bibfnamefont {A.~S.}\ \bibnamefont
  {Brichkin}}, \bibinfo {author} {\bibfnamefont {S.~I.}\ \bibnamefont
  {Novikov}}, \bibinfo {author} {\bibfnamefont {S.}~\bibnamefont
  {H{\"o}fling}}, \bibinfo {author} {\bibfnamefont {C.}~\bibnamefont
  {Schneider}}, \bibinfo {author} {\bibfnamefont {M.}~\bibnamefont {Kamp}},
  \bibinfo {author} {\bibfnamefont {A.}~\bibnamefont {Forchel}},\ and\ \bibinfo
  {author} {\bibfnamefont {V.~D.}\ \bibnamefont {Kulakovskii}},\ }\bibfield
  {title} {\bibinfo {title} {{Nonlinear route to intrinsic Josephson
  oscillations in spinor cavity-polariton condensates}},\ }\href
  {https://doi.org/10.1103/PhysRevB.90.235309} {\bibfield  {journal} {\bibinfo
  {journal} {Phys. Rev. B}\ }\textbf {\bibinfo {volume} {90}},\ \bibinfo
  {pages} {235309} (\bibinfo {year} {2014})}\BibitemShut {NoStop}%
\bibitem [{\citenamefont {Pigeon}\ \emph {et~al.}(2011)\citenamefont {Pigeon},
  \citenamefont {Carusotto},\ and\ \citenamefont {Ciuti}}]{Pigeon2011}%
  \BibitemOpen
  \bibfield  {author} {\bibinfo {author} {\bibfnamefont {S.}~\bibnamefont
  {Pigeon}}, \bibinfo {author} {\bibfnamefont {I.}~\bibnamefont {Carusotto}},\
  and\ \bibinfo {author} {\bibfnamefont {C.}~\bibnamefont {Ciuti}},\ }\bibfield
   {title} {\bibinfo {title} {{Hydrodynamic nucleation of vortices and solitons
  in a resonantly excited polariton superfluid}},\ }\href
  {https://doi.org/10.1103/PhysRevB.83.144513} {\bibfield  {journal} {\bibinfo
  {journal} {Phys. Rev. B}\ }\textbf {\bibinfo {volume} {83}},\ \bibinfo
  {pages} {144513} (\bibinfo {year} {2011})}\BibitemShut {NoStop}%
\bibitem [{\citenamefont {Gavrilov}(2020{\natexlab{b}})}]{Gavrilov2020}%
  \BibitemOpen
  \bibfield  {author} {\bibinfo {author} {\bibfnamefont {S.~S.}\ \bibnamefont
  {Gavrilov}},\ }\bibfield  {title} {\bibinfo {title} {{Spontaneous formation
  of vortices and gray solitons in a spinor polariton fluid under coherent
  driving}},\ }\href {https://doi.org/10.1103/PhysRevB.102.104307} {\bibfield
  {journal} {\bibinfo  {journal} {Phys. Rev. B}\ }\textbf {\bibinfo {volume}
  {102}},\ \bibinfo {pages} {104307} (\bibinfo {year}
  {2020}{\natexlab{b}})}\BibitemShut {NoStop}%
\bibitem [{\citenamefont {Ihara}\ and\ \citenamefont
  {Kasamatsu}(2019)}]{Ihara2019}%
  \BibitemOpen
  \bibfield  {author} {\bibinfo {author} {\bibfnamefont {K.}~\bibnamefont
  {Ihara}}\ and\ \bibinfo {author} {\bibfnamefont {K.}~\bibnamefont
  {Kasamatsu}},\ }\bibfield  {title} {\bibinfo {title} {{Transverse instability
  and disintegration of a domain wall of a relative phase in coherently coupled
  two-component Bose-Einstein condensates}},\ }\href
  {https://doi.org/10.1103/PhysRevA.100.013630} {\bibfield  {journal} {\bibinfo
   {journal} {Phys. Rev. A}\ }\textbf {\bibinfo {volume} {100}},\ \bibinfo
  {pages} {013630} (\bibinfo {year} {2019})}\BibitemShut {NoStop}%
\bibitem [{\citenamefont {Gallem{\'\i}}\ \emph {et~al.}(2019)\citenamefont
  {Gallem{\'\i}}, \citenamefont {Pitaevskii}, \citenamefont {Stringari},\ and\
  \citenamefont {Recati}}]{Gallemi2019}%
  \BibitemOpen
  \bibfield  {author} {\bibinfo {author} {\bibfnamefont {A.}~\bibnamefont
  {Gallem{\'\i}}}, \bibinfo {author} {\bibfnamefont {L.~P.}\ \bibnamefont
  {Pitaevskii}}, \bibinfo {author} {\bibfnamefont {S.}~\bibnamefont
  {Stringari}},\ and\ \bibinfo {author} {\bibfnamefont {A.}~\bibnamefont
  {Recati}},\ }\bibfield  {title} {\bibinfo {title} {{Decay of the relative
  phase domain wall into confined vortex pairs: The case of a coherently
  coupled bosonic mixture}},\ }\href
  {https://doi.org/10.1103/PhysRevA.100.023607} {\bibfield  {journal} {\bibinfo
   {journal} {Phys. Rev. A}\ }\textbf {\bibinfo {volume} {100}},\ \bibinfo
  {pages} {023607} (\bibinfo {year} {2019})}\BibitemShut {NoStop}%
\bibitem [{\citenamefont {Gallem{\'\i}}\ \emph {et~al.}(2016)\citenamefont
  {Gallem{\'\i}}, \citenamefont {Guilleumas}, \citenamefont {Mayol},\ and\
  \citenamefont {Mateo}}]{Gallemi2016}%
  \BibitemOpen
  \bibfield  {author} {\bibinfo {author} {\bibfnamefont {A.}~\bibnamefont
  {Gallem{\'\i}}}, \bibinfo {author} {\bibfnamefont {M.}~\bibnamefont
  {Guilleumas}}, \bibinfo {author} {\bibfnamefont {R.}~\bibnamefont {Mayol}},\
  and\ \bibinfo {author} {\bibfnamefont {A.~M.}\ \bibnamefont {Mateo}},\
  }\bibfield  {title} {\bibinfo {title} {{Multidimensional Josephson vortices
  in spin-orbit-coupled Bose-Einstein condensates: Snake instability and decay
  through vortex dipoles}},\ }\href
  {https://doi.org/10.1103/PhysRevA.93.033618} {\bibfield  {journal} {\bibinfo
  {journal} {Phys. Rev. A}\ }\textbf {\bibinfo {volume} {93}},\ \bibinfo
  {pages} {033618} (\bibinfo {year} {2016})}\BibitemShut {NoStop}%
\bibitem [{\citenamefont {Qu}\ \emph {et~al.}(2017)\citenamefont {Qu},
  \citenamefont {Tylutki}, \citenamefont {Stringari},\ and\ \citenamefont
  {Pitaevskii}}]{Qu2017}%
  \BibitemOpen
  \bibfield  {author} {\bibinfo {author} {\bibfnamefont {C.}~\bibnamefont
  {Qu}}, \bibinfo {author} {\bibfnamefont {M.}~\bibnamefont {Tylutki}},
  \bibinfo {author} {\bibfnamefont {S.}~\bibnamefont {Stringari}},\ and\
  \bibinfo {author} {\bibfnamefont {L.~P.}\ \bibnamefont {Pitaevskii}},\
  }\bibfield  {title} {\bibinfo {title} {{Magnetic solitons in Rabi-coupled
  Bose-Einstein condensates}},\ }\href
  {https://doi.org/10.1103/PhysRevA.95.033614} {\bibfield  {journal} {\bibinfo
  {journal} {Phys. Rev. A}\ }\textbf {\bibinfo {volume} {95}},\ \bibinfo
  {pages} {033614} (\bibinfo {year} {2017})}\BibitemShut {NoStop}%
\bibitem [{\citenamefont {Yu}\ and\ \citenamefont {Blakie}(2021)}]{Yu2021}%
  \BibitemOpen
  \bibfield  {author} {\bibinfo {author} {\bibfnamefont {X.}~\bibnamefont
  {Yu}}\ and\ \bibinfo {author} {\bibfnamefont {P.~B.}\ \bibnamefont
  {Blakie}},\ }\bibfield  {title} {\bibinfo {title} {{Dark-soliton-like
  magnetic domain walls in a two-dimensional ferromagnetic superfluid}},\
  }\href {https://doi.org/10.1103/PhysRevResearch.3.023043} {\bibfield
  {journal} {\bibinfo  {journal} {Phys. Rev. Res.}\ }\textbf {\bibinfo {volume}
  {3}},\ \bibinfo {pages} {023043} (\bibinfo {year} {2021})}\BibitemShut
  {NoStop}%
\bibitem [{\citenamefont {Yu}\ and\ \citenamefont {Blakie}(2022)}]{Yu2022}%
  \BibitemOpen
  \bibfield  {author} {\bibinfo {author} {\bibfnamefont {X.}~\bibnamefont
  {Yu}}\ and\ \bibinfo {author} {\bibfnamefont {P.~B.}\ \bibnamefont
  {Blakie}},\ }\bibfield  {title} {\bibinfo {title} {{Propagating Ferrodark
  Solitons in a Superfluid: Exact Solutions and Anomalous Dynamics}},\ }\href
  {https://doi.org/10.1103/PhysRevLett.128.125301} {\bibfield  {journal}
  {\bibinfo  {journal} {Phys. Rev. Lett.}\ }\textbf {\bibinfo {volume} {128}},\
  \bibinfo {pages} {125301} (\bibinfo {year} {2022})}\BibitemShut {NoStop}%
\bibitem [{\citenamefont {Pitaevskii}\ and\ \citenamefont
  {Stringari}(2016)}]{Pitaevskii.book}%
  \BibitemOpen
  \bibfield  {author} {\bibinfo {author} {\bibfnamefont {L.}~\bibnamefont
  {Pitaevskii}}\ and\ \bibinfo {author} {\bibfnamefont {S.}~\bibnamefont
  {Stringari}},\ }\href@noop {} {\emph {\bibinfo {title} {{Bose-Einstein
  Condensation and Superfluidity}}}}\ (\bibinfo  {publisher} {Oxford University
  Press},\ \bibinfo {address} {New York},\ \bibinfo {year} {2016})\BibitemShut
  {NoStop}%
\bibitem [{\citenamefont
  {Gavrilov}(2020{\natexlab{c}})}]{Gavrilov2020.usp.ru.en}%
  \BibitemOpen
  \bibfield  {author} {\bibinfo {author} {\bibfnamefont {S.~S.}\ \bibnamefont
  {Gavrilov}},\ }\bibfield  {title} {\bibinfo {title} {{Nonequilibrium
  transitions, chaos, and chimera states in exciton-polariton systems}},\
  }\href {https://doi.org/10.3367/UFNr.2019.04.038549} {\bibfield  {journal}
  {\bibinfo  {journal} {Usp. Fiz. Nauk}\ }\textbf {\bibinfo {volume} {190}},\
  \bibinfo {pages} {137} (\bibinfo {year} {2020}{\natexlab{c}})}\BibitemShut
  {NoStop}%
\bibitem [{\citenamefont {Rubakov}(2002)}]{Rubakov.book.CTGF.en}%
  \BibitemOpen
  \bibfield  {author} {\bibinfo {author} {\bibfnamefont {V.}~\bibnamefont
  {Rubakov}},\ }\href@noop {} {\emph {\bibinfo {title} {{Classical Theory of
  Gauge Fields}}}}\ (\bibinfo  {publisher} {Princeton University Press},\
  \bibinfo {year} {2002})\BibitemShut {NoStop}%
\bibitem [{\citenamefont {Boulier}\ \emph {et~al.}(2015)\citenamefont
  {Boulier}, \citenamefont {Ter{\c c}as}, \citenamefont {Solnyshkov},
  \citenamefont {Glorieux}, \citenamefont {Giacobino}, \citenamefont
  {Malpuech},\ and\ \citenamefont {Bramati}}]{Boulier2015}%
  \BibitemOpen
  \bibfield  {author} {\bibinfo {author} {\bibfnamefont {T.}~\bibnamefont
  {Boulier}}, \bibinfo {author} {\bibfnamefont {H.}~\bibnamefont {Ter{\c
  c}as}}, \bibinfo {author} {\bibfnamefont {D.~D.}\ \bibnamefont {Solnyshkov}},
  \bibinfo {author} {\bibfnamefont {Q.}~\bibnamefont {Glorieux}}, \bibinfo
  {author} {\bibfnamefont {E.}~\bibnamefont {Giacobino}}, \bibinfo {author}
  {\bibfnamefont {G.}~\bibnamefont {Malpuech}},\ and\ \bibinfo {author}
  {\bibfnamefont {A.}~\bibnamefont {Bramati}},\ }\bibfield  {title} {\bibinfo
  {title} {{Vortex Chain in a Resonantly Pumped Polariton Superfluid}},\ }\href
  {https://doi.org/10.1038/srep09230} {\bibfield  {journal} {\bibinfo
  {journal} {Sci. Rep.}\ }\textbf {\bibinfo {volume} {5}},\ \bibinfo {pages}
  {9230} (\bibinfo {year} {2015})}\BibitemShut {NoStop}%
\bibitem [{\citenamefont {Dominici}\ \emph {et~al.}(2015)\citenamefont
  {Dominici}, \citenamefont {Dagvadorj}, \citenamefont {Fellows}, \citenamefont
  {Ballarini}, \citenamefont {{De Giorgi}}, \citenamefont {Marchetti},
  \citenamefont {Piccirillo}, \citenamefont {Marrucci}, \citenamefont
  {Bramati}, \citenamefont {Gigli}, \citenamefont {Szyma{\'n}ska},\ and\
  \citenamefont {Sanvitto}}]{Dominici2015}%
  \BibitemOpen
  \bibfield  {author} {\bibinfo {author} {\bibfnamefont {L.}~\bibnamefont
  {Dominici}}, \bibinfo {author} {\bibfnamefont {G.}~\bibnamefont {Dagvadorj}},
  \bibinfo {author} {\bibfnamefont {J.~M.}\ \bibnamefont {Fellows}}, \bibinfo
  {author} {\bibfnamefont {D.}~\bibnamefont {Ballarini}}, \bibinfo {author}
  {\bibfnamefont {M.}~\bibnamefont {{De Giorgi}}}, \bibinfo {author}
  {\bibfnamefont {F.~M.}\ \bibnamefont {Marchetti}}, \bibinfo {author}
  {\bibfnamefont {B.}~\bibnamefont {Piccirillo}}, \bibinfo {author}
  {\bibfnamefont {L.}~\bibnamefont {Marrucci}}, \bibinfo {author}
  {\bibfnamefont {A.}~\bibnamefont {Bramati}}, \bibinfo {author} {\bibfnamefont
  {G.}~\bibnamefont {Gigli}}, \bibinfo {author} {\bibfnamefont {M.~H.}\
  \bibnamefont {Szyma{\'n}ska}},\ and\ \bibinfo {author} {\bibfnamefont
  {D.}~\bibnamefont {Sanvitto}},\ }\bibfield  {title} {\bibinfo {title}
  {{Vortex and half-vortex dynamics in a nonlinear spinor quantum fluid}},\
  }\href {https://doi.org/10.1126/sciadv.1500807} {\bibfield  {journal}
  {\bibinfo  {journal} {Sci. Adv.}\ }\textbf {\bibinfo {volume} {1}},\ \bibinfo
  {pages} {e1500807} (\bibinfo {year} {2015})}\BibitemShut {NoStop}%
\bibitem [{\citenamefont {Dominici}\ \emph {et~al.}(2018)\citenamefont
  {Dominici}, \citenamefont {Carretero-Gonz{\'a}lez}, \citenamefont
  {Gianfrate}, \citenamefont {Cuevas-Maraver}, \citenamefont {Rodrigues},
  \citenamefont {Frantzeskakis}, \citenamefont {Lerario}, \citenamefont
  {Ballarini}, \citenamefont {{De Giorgi}}, \citenamefont {Gigli},
  \citenamefont {Kevrekidis},\ and\ \citenamefont {Sanvitto}}]{Dominici2018}%
  \BibitemOpen
  \bibfield  {author} {\bibinfo {author} {\bibfnamefont {L.}~\bibnamefont
  {Dominici}}, \bibinfo {author} {\bibfnamefont {R.}~\bibnamefont
  {Carretero-Gonz{\'a}lez}}, \bibinfo {author} {\bibfnamefont {A.}~\bibnamefont
  {Gianfrate}}, \bibinfo {author} {\bibfnamefont {J.}~\bibnamefont
  {Cuevas-Maraver}}, \bibinfo {author} {\bibfnamefont {A.~S.}\ \bibnamefont
  {Rodrigues}}, \bibinfo {author} {\bibfnamefont {D.~J.}\ \bibnamefont
  {Frantzeskakis}}, \bibinfo {author} {\bibfnamefont {G.}~\bibnamefont
  {Lerario}}, \bibinfo {author} {\bibfnamefont {D.}~\bibnamefont {Ballarini}},
  \bibinfo {author} {\bibfnamefont {M.}~\bibnamefont {{De Giorgi}}}, \bibinfo
  {author} {\bibfnamefont {G.}~\bibnamefont {Gigli}}, \bibinfo {author}
  {\bibfnamefont {P.~G.}\ \bibnamefont {Kevrekidis}},\ and\ \bibinfo {author}
  {\bibfnamefont {D.}~\bibnamefont {Sanvitto}},\ }\bibfield  {title} {\bibinfo
  {title} {{Interactions and scattering of quantum vortices in a polariton
  fluid}},\ }\href {https://doi.org/10.1038/s41467-018-03736-5} {\bibfield
  {journal} {\bibinfo  {journal} {Nat. Commun.}\ }\textbf {\bibinfo {volume}
  {9}},\ \bibinfo {pages} {1467} (\bibinfo {year} {2018})}\BibitemShut
  {NoStop}%
\bibitem [{\citenamefont {Carusotto}\ and\ \citenamefont
  {Ciuti}(2013)}]{Carusotto2013}%
  \BibitemOpen
  \bibfield  {author} {\bibinfo {author} {\bibfnamefont {I.}~\bibnamefont
  {Carusotto}}\ and\ \bibinfo {author} {\bibfnamefont {C.}~\bibnamefont
  {Ciuti}},\ }\bibfield  {title} {\bibinfo {title} {{Quantum fluids of
  light}},\ }\href {https://doi.org/10.1103/RevModPhys.85.299} {\bibfield
  {journal} {\bibinfo  {journal} {Rev. Mod. Phys.}\ }\textbf {\bibinfo {volume}
  {85}},\ \bibinfo {pages} {299} (\bibinfo {year} {2013})}\BibitemShut
  {NoStop}%
\bibitem [{\citenamefont {Amo}\ \emph {et~al.}(2011)\citenamefont {Amo},
  \citenamefont {Pigeon}, \citenamefont {Sanvitto}, \citenamefont {Sala},
  \citenamefont {Hivet}, \citenamefont {Carusotto}, \citenamefont {Pisanello},
  \citenamefont {Lem{\'e}nager}, \citenamefont {Houdr{\'e}}, \citenamefont
  {Giacobino}, \citenamefont {Ciuti},\ and\ \citenamefont {Bramati}}]{Amo2011}%
  \BibitemOpen
  \bibfield  {author} {\bibinfo {author} {\bibfnamefont {A.}~\bibnamefont
  {Amo}}, \bibinfo {author} {\bibfnamefont {S.}~\bibnamefont {Pigeon}},
  \bibinfo {author} {\bibfnamefont {D.}~\bibnamefont {Sanvitto}}, \bibinfo
  {author} {\bibfnamefont {V.~G.}\ \bibnamefont {Sala}}, \bibinfo {author}
  {\bibfnamefont {R.}~\bibnamefont {Hivet}}, \bibinfo {author} {\bibfnamefont
  {I.}~\bibnamefont {Carusotto}}, \bibinfo {author} {\bibfnamefont
  {F.}~\bibnamefont {Pisanello}}, \bibinfo {author} {\bibfnamefont
  {G.}~\bibnamefont {Lem{\'e}nager}}, \bibinfo {author} {\bibfnamefont
  {R.}~\bibnamefont {Houdr{\'e}}}, \bibinfo {author} {\bibfnamefont
  {E.}~\bibnamefont {Giacobino}}, \bibinfo {author} {\bibfnamefont
  {C.}~\bibnamefont {Ciuti}},\ and\ \bibinfo {author} {\bibfnamefont
  {A.}~\bibnamefont {Bramati}},\ }\bibfield  {title} {\bibinfo {title}
  {{Polariton Superfluids Reveal Quantum Hydrodynamic Solitons}},\ }\href
  {https://doi.org/10.1126/science.1202307} {\bibfield  {journal} {\bibinfo
  {journal} {Science}\ }\textbf {\bibinfo {volume} {332}},\ \bibinfo {pages}
  {1167} (\bibinfo {year} {2011})}\BibitemShut {NoStop}%
\bibitem [{\citenamefont {Pigeon}\ and\ \citenamefont
  {Bramati}(2017)}]{Pigeon2017}%
  \BibitemOpen
  \bibfield  {author} {\bibinfo {author} {\bibfnamefont {S.}~\bibnamefont
  {Pigeon}}\ and\ \bibinfo {author} {\bibfnamefont {A.}~\bibnamefont
  {Bramati}},\ }\bibfield  {title} {\bibinfo {title} {{Sustained propagation
  and control of topological excitations in polariton superfluid}},\ }\href
  {https://doi.org/10.1088/1367-2630/aa849c} {\bibfield  {journal} {\bibinfo
  {journal} {New J. Phys.}\ }\textbf {\bibinfo {volume} {19}},\ \bibinfo
  {pages} {095004} (\bibinfo {year} {2017})}\BibitemShut {NoStop}%
\bibitem [{\citenamefont {Whittaker}(2007)}]{Whittaker2007}%
  \BibitemOpen
  \bibfield  {author} {\bibinfo {author} {\bibfnamefont {D.~M.}\ \bibnamefont
  {Whittaker}},\ }\bibfield  {title} {\bibinfo {title} {{Vortices in the
  microcavity optical parametric oscillator}},\ }\href
  {https://doi.org/10.1016/j.spmi.2007.03.032} {\bibfield  {journal} {\bibinfo
  {journal} {Superlattices Microstruct.}\ }\textbf {\bibinfo {volume} {41}},\
  \bibinfo {pages} {297} (\bibinfo {year} {2007})}\BibitemShut {NoStop}%
\bibitem [{\citenamefont {Krizhanovskii}\ \emph {et~al.}(2010)\citenamefont
  {Krizhanovskii}, \citenamefont {Whittaker}, \citenamefont {Bradley},
  \citenamefont {Guda}, \citenamefont {Sarkar}, \citenamefont {Sanvitto},
  \citenamefont {Vi{\~n}a}, \citenamefont {Cerda}, \citenamefont {Santos},
  \citenamefont {Biermann}, \citenamefont {Hey},\ and\ \citenamefont
  {Skolnick}}]{Krizhanovskii2010}%
  \BibitemOpen
  \bibfield  {author} {\bibinfo {author} {\bibfnamefont {D.~N.}\ \bibnamefont
  {Krizhanovskii}}, \bibinfo {author} {\bibfnamefont {D.~M.}\ \bibnamefont
  {Whittaker}}, \bibinfo {author} {\bibfnamefont {R.~A.}\ \bibnamefont
  {Bradley}}, \bibinfo {author} {\bibfnamefont {K.}~\bibnamefont {Guda}},
  \bibinfo {author} {\bibfnamefont {D.}~\bibnamefont {Sarkar}}, \bibinfo
  {author} {\bibfnamefont {D.}~\bibnamefont {Sanvitto}}, \bibinfo {author}
  {\bibfnamefont {L.}~\bibnamefont {Vi{\~n}a}}, \bibinfo {author}
  {\bibfnamefont {E.}~\bibnamefont {Cerda}}, \bibinfo {author} {\bibfnamefont
  {P.}~\bibnamefont {Santos}}, \bibinfo {author} {\bibfnamefont
  {K.}~\bibnamefont {Biermann}}, \bibinfo {author} {\bibfnamefont
  {R.}~\bibnamefont {Hey}},\ and\ \bibinfo {author} {\bibfnamefont {M.~S.}\
  \bibnamefont {Skolnick}},\ }\bibfield  {title} {\bibinfo {title} {{Effect of
  Interactions on Vortices in a Nonequilibrium Polariton Condensate}},\ }\href
  {https://doi.org/10.1103/PhysRevLett.104.126402} {\bibfield  {journal}
  {\bibinfo  {journal} {Phys. Rev. Lett.}\ }\textbf {\bibinfo {volume} {104}},\
  \bibinfo {pages} {126402} (\bibinfo {year} {2010})}\BibitemShut {NoStop}%
\bibitem [{\citenamefont {Marchetti}\ \emph {et~al.}(2010)\citenamefont
  {Marchetti}, \citenamefont {Szyma{\'n}ska}, \citenamefont {Tejedor},\ and\
  \citenamefont {Whittaker}}]{Marchetti2010}%
  \BibitemOpen
  \bibfield  {author} {\bibinfo {author} {\bibfnamefont {F.~M.}\ \bibnamefont
  {Marchetti}}, \bibinfo {author} {\bibfnamefont {M.~H.}\ \bibnamefont
  {Szyma{\'n}ska}}, \bibinfo {author} {\bibfnamefont {C.}~\bibnamefont
  {Tejedor}},\ and\ \bibinfo {author} {\bibfnamefont {D.~M.}\ \bibnamefont
  {Whittaker}},\ }\bibfield  {title} {\bibinfo {title} {{Spontaneous and
  Triggered Vortices in Polariton Optical-Parametric-Oscillator Superfluids}},\
  }\href {https://doi.org/10.1103/PhysRevLett.105.063902} {\bibfield  {journal}
  {\bibinfo  {journal} {Phys. Rev. Lett.}\ }\textbf {\bibinfo {volume} {105}},\
  \bibinfo {pages} {063902} (\bibinfo {year} {2010})}\BibitemShut {NoStop}%
\bibitem [{\citenamefont {Sanvitto}\ \emph {et~al.}(2010)\citenamefont
  {Sanvitto}, \citenamefont {Marchetti}, \citenamefont {Szyma{\'n}ska},
  \citenamefont {Tosi}, \citenamefont {Baudisch}, \citenamefont {Laussy},
  \citenamefont {Krizhanovskii}, \citenamefont {Skolnick}, \citenamefont
  {Marrucci}, \citenamefont {Lema{\^\i}{}tre}, \citenamefont {Bloch},
  \citenamefont {Tejedor},\ and\ \citenamefont {Vi{\~n}a}}]{Sanvitto2010}%
  \BibitemOpen
  \bibfield  {author} {\bibinfo {author} {\bibfnamefont {D.}~\bibnamefont
  {Sanvitto}}, \bibinfo {author} {\bibfnamefont {F.~M.}\ \bibnamefont
  {Marchetti}}, \bibinfo {author} {\bibfnamefont {M.~H.}\ \bibnamefont
  {Szyma{\'n}ska}}, \bibinfo {author} {\bibfnamefont {G.}~\bibnamefont {Tosi}},
  \bibinfo {author} {\bibfnamefont {M.}~\bibnamefont {Baudisch}}, \bibinfo
  {author} {\bibfnamefont {F.~P.}\ \bibnamefont {Laussy}}, \bibinfo {author}
  {\bibfnamefont {D.~N.}\ \bibnamefont {Krizhanovskii}}, \bibinfo {author}
  {\bibfnamefont {M.~S.}\ \bibnamefont {Skolnick}}, \bibinfo {author}
  {\bibfnamefont {L.}~\bibnamefont {Marrucci}}, \bibinfo {author}
  {\bibfnamefont {A.}~\bibnamefont {Lema{\^\i}{}tre}}, \bibinfo {author}
  {\bibfnamefont {J.}~\bibnamefont {Bloch}}, \bibinfo {author} {\bibfnamefont
  {C.}~\bibnamefont {Tejedor}},\ and\ \bibinfo {author} {\bibfnamefont
  {L.}~\bibnamefont {Vi{\~n}a}},\ }\bibfield  {title} {\bibinfo {title}
  {{Persistent currents and quantized vortices in a polariton superfluid}},\
  }\href {https://doi.org/10.1038/nphys1668} {\bibfield  {journal} {\bibinfo
  {journal} {Nat. Phys.}\ }\textbf {\bibinfo {volume} {6}},\ \bibinfo {pages}
  {527} (\bibinfo {year} {2010})}\BibitemShut {NoStop}%
\bibitem [{\citenamefont {Dagvadorj}\ \emph {et~al.}(2015)\citenamefont
  {Dagvadorj}, \citenamefont {Fellows}, \citenamefont {{Matyja\ifmmode
  \acute{s}\else {\'s}\fi{}kiewicz}}, \citenamefont {Marchetti}, \citenamefont
  {Carusotto},\ and\ \citenamefont {{Szyma\ifmmode \acute{n}\else
  {\'n}\fi{}ska}}}]{Dagvadorj2015}%
  \BibitemOpen
  \bibfield  {author} {\bibinfo {author} {\bibfnamefont {G.}~\bibnamefont
  {Dagvadorj}}, \bibinfo {author} {\bibfnamefont {J.~M.}\ \bibnamefont
  {Fellows}}, \bibinfo {author} {\bibfnamefont {S.}~\bibnamefont
  {{Matyja\ifmmode \acute{s}\else {\'s}\fi{}kiewicz}}}, \bibinfo {author}
  {\bibfnamefont {F.~M.}\ \bibnamefont {Marchetti}}, \bibinfo {author}
  {\bibfnamefont {I.}~\bibnamefont {Carusotto}},\ and\ \bibinfo {author}
  {\bibfnamefont {M.~H.}\ \bibnamefont {{Szyma\ifmmode \acute{n}\else
  {\'n}\fi{}ska}}},\ }\bibfield  {title} {\bibinfo {title} {{Nonequilibrium
  Phase Transition in a Two-Dimensional Driven Open Quantum System}},\ }\href
  {https://doi.org/10.1103/PhysRevX.5.041028} {\bibfield  {journal} {\bibinfo
  {journal} {Phys. Rev. X}\ }\textbf {\bibinfo {volume} {5}},\ \bibinfo {pages}
  {041028} (\bibinfo {year} {2015})}\BibitemShut {NoStop}%
\bibitem [{\citenamefont {Koniakhin}\ \emph {et~al.}(2019)\citenamefont
  {Koniakhin}, \citenamefont {Bleu}, \citenamefont {Stupin}, \citenamefont
  {Pigeon}, \citenamefont {Maitre}, \citenamefont {Claude}, \citenamefont
  {Lerario}, \citenamefont {Glorieux}, \citenamefont {Bramati}, \citenamefont
  {Solnyshkov},\ and\ \citenamefont {Malpuech}}]{Koniakhin2019}%
  \BibitemOpen
  \bibfield  {author} {\bibinfo {author} {\bibfnamefont {S.~V.}\ \bibnamefont
  {Koniakhin}}, \bibinfo {author} {\bibfnamefont {O.}~\bibnamefont {Bleu}},
  \bibinfo {author} {\bibfnamefont {D.~D.}\ \bibnamefont {Stupin}}, \bibinfo
  {author} {\bibfnamefont {S.}~\bibnamefont {Pigeon}}, \bibinfo {author}
  {\bibfnamefont {A.}~\bibnamefont {Maitre}}, \bibinfo {author} {\bibfnamefont
  {F.}~\bibnamefont {Claude}}, \bibinfo {author} {\bibfnamefont
  {G.}~\bibnamefont {Lerario}}, \bibinfo {author} {\bibfnamefont
  {Q.}~\bibnamefont {Glorieux}}, \bibinfo {author} {\bibfnamefont
  {A.}~\bibnamefont {Bramati}}, \bibinfo {author} {\bibfnamefont
  {D.}~\bibnamefont {Solnyshkov}},\ and\ \bibinfo {author} {\bibfnamefont
  {G.}~\bibnamefont {Malpuech}},\ }\bibfield  {title} {\bibinfo {title}
  {{Stationary Quantum Vortex Street in a Driven-Dissipative Quantum Fluid of
  Light}},\ }\href {https://doi.org/10.1103/PhysRevLett.123.215301} {\bibfield
  {journal} {\bibinfo  {journal} {Phys. Rev. Lett.}\ }\textbf {\bibinfo
  {volume} {123}},\ \bibinfo {pages} {215301} (\bibinfo {year}
  {2019})}\BibitemShut {NoStop}%
\bibitem [{\citenamefont {Claude}\ \emph {et~al.}(2020)\citenamefont {Claude},
  \citenamefont {Koniakhin}, \citenamefont {Ma{\^\i}tre}, \citenamefont
  {Pigeon}, \citenamefont {Lerario}, \citenamefont {Stupin}, \citenamefont
  {Glorieux}, \citenamefont {Giacobino}, \citenamefont {Solnyshkov},
  \citenamefont {Malpuech},\ and\ \citenamefont {Bramati}}]{Claude2020}%
  \BibitemOpen
  \bibfield  {author} {\bibinfo {author} {\bibfnamefont {F.}~\bibnamefont
  {Claude}}, \bibinfo {author} {\bibfnamefont {S.~V.}\ \bibnamefont
  {Koniakhin}}, \bibinfo {author} {\bibfnamefont {A.}~\bibnamefont
  {Ma{\^\i}tre}}, \bibinfo {author} {\bibfnamefont {S.}~\bibnamefont {Pigeon}},
  \bibinfo {author} {\bibfnamefont {G.}~\bibnamefont {Lerario}}, \bibinfo
  {author} {\bibfnamefont {D.~D.}\ \bibnamefont {Stupin}}, \bibinfo {author}
  {\bibfnamefont {Q.}~\bibnamefont {Glorieux}}, \bibinfo {author}
  {\bibfnamefont {E.}~\bibnamefont {Giacobino}}, \bibinfo {author}
  {\bibfnamefont {D.}~\bibnamefont {Solnyshkov}}, \bibinfo {author}
  {\bibfnamefont {G.}~\bibnamefont {Malpuech}},\ and\ \bibinfo {author}
  {\bibfnamefont {A.}~\bibnamefont {Bramati}},\ }\bibfield  {title} {\bibinfo
  {title} {{Taming the snake instabilities in a polariton superfluid}},\ }\href
  {https://doi.org/10.1364/OPTICA.405946} {\bibfield  {journal} {\bibinfo
  {journal} {Optica}\ }\textbf {\bibinfo {volume} {7}},\ \bibinfo {pages}
  {1660} (\bibinfo {year} {2020})}\BibitemShut {NoStop}%
\bibitem [{\citenamefont {Kierzenka}\ and\ \citenamefont
  {Shampine}(2001)}]{Kierzenka2001}%
  \BibitemOpen
  \bibfield  {author} {\bibinfo {author} {\bibfnamefont {J.}~\bibnamefont
  {Kierzenka}}\ and\ \bibinfo {author} {\bibfnamefont {L.~F.}\ \bibnamefont
  {Shampine}},\ }\bibfield  {title} {\bibinfo {title} {{A BVP solver based on
  residual control and the Maltab PSE}},\ }\href
  {https://doi.org/10.1145/502800.502801} {\bibfield  {journal} {\bibinfo
  {journal} {ACM Trans. Math. Softw.}\ }\textbf {\bibinfo {volume} {27}},\
  \bibinfo {pages} {299} (\bibinfo {year} {2001})}\BibitemShut {NoStop}%
\bibitem [{\citenamefont {Virtanen}\ \emph {et~al.}(2020)\citenamefont
  {Virtanen}, \citenamefont {Gommers}, \citenamefont {Oliphant}, \citenamefont
  {Haberland}, \citenamefont {Reddy}, \citenamefont {Cournapeau}, \citenamefont
  {Burovski}, \citenamefont {Peterson}, \citenamefont {Weckesser},
  \citenamefont {Bright}, \citenamefont {{van der Walt}}, \citenamefont
  {Brett}, \citenamefont {Wilson}, \citenamefont {Millman}, \citenamefont
  {Mayorov}, \citenamefont {Nelson}, \citenamefont {Jones}, \citenamefont
  {Kern}, \citenamefont {Larson}, \citenamefont {Carey}, \citenamefont {Polat},
  \citenamefont {Feng}, \citenamefont {Moore}, \citenamefont {{VanderPlas}},
  \citenamefont {Laxalde}, \citenamefont {Perktold}, \citenamefont {Cimrman},
  \citenamefont {Henriksen}, \citenamefont {Quintero}, \citenamefont {Harris},
  \citenamefont {Archibald}, \citenamefont {Ribeiro}, \citenamefont
  {Pedregosa}, \citenamefont {{van Mulbregt}},\ and\ \citenamefont {{SciPy 1.0
  Contributors}}}]{Virtanen2020}%
  \BibitemOpen
  \bibfield  {author} {\bibinfo {author} {\bibfnamefont {P.}~\bibnamefont
  {Virtanen}}, \bibinfo {author} {\bibfnamefont {R.}~\bibnamefont {Gommers}},
  \bibinfo {author} {\bibfnamefont {T.~E.}\ \bibnamefont {Oliphant}}, \bibinfo
  {author} {\bibfnamefont {M.}~\bibnamefont {Haberland}}, \bibinfo {author}
  {\bibfnamefont {T.}~\bibnamefont {Reddy}}, \bibinfo {author} {\bibfnamefont
  {D.}~\bibnamefont {Cournapeau}}, \bibinfo {author} {\bibfnamefont
  {E.}~\bibnamefont {Burovski}}, \bibinfo {author} {\bibfnamefont
  {P.}~\bibnamefont {Peterson}}, \bibinfo {author} {\bibfnamefont
  {W.}~\bibnamefont {Weckesser}}, \bibinfo {author} {\bibfnamefont
  {J.}~\bibnamefont {Bright}}, \bibinfo {author} {\bibfnamefont {S.~J.}\
  \bibnamefont {{van der Walt}}}, \bibinfo {author} {\bibfnamefont
  {M.}~\bibnamefont {Brett}}, \bibinfo {author} {\bibfnamefont
  {J.}~\bibnamefont {Wilson}}, \bibinfo {author} {\bibfnamefont {K.~J.}\
  \bibnamefont {Millman}}, \bibinfo {author} {\bibfnamefont {N.}~\bibnamefont
  {Mayorov}}, \bibinfo {author} {\bibfnamefont {A.~R.~J.}\ \bibnamefont
  {Nelson}}, \bibinfo {author} {\bibfnamefont {E.}~\bibnamefont {Jones}},
  \bibinfo {author} {\bibfnamefont {R.}~\bibnamefont {Kern}}, \bibinfo {author}
  {\bibfnamefont {E.}~\bibnamefont {Larson}}, \bibinfo {author} {\bibfnamefont
  {C.~J.}\ \bibnamefont {Carey}}, \bibinfo {author} {\bibfnamefont
  {{\.I}.}~\bibnamefont {Polat}}, \bibinfo {author} {\bibfnamefont
  {Y.}~\bibnamefont {Feng}}, \bibinfo {author} {\bibfnamefont {E.~W.}\
  \bibnamefont {Moore}}, \bibinfo {author} {\bibfnamefont {J.}~\bibnamefont
  {{VanderPlas}}}, \bibinfo {author} {\bibfnamefont {D.}~\bibnamefont
  {Laxalde}}, \bibinfo {author} {\bibfnamefont {J.}~\bibnamefont {Perktold}},
  \bibinfo {author} {\bibfnamefont {R.}~\bibnamefont {Cimrman}}, \bibinfo
  {author} {\bibfnamefont {I.}~\bibnamefont {Henriksen}}, \bibinfo {author}
  {\bibfnamefont {E.~A.}\ \bibnamefont {Quintero}}, \bibinfo {author}
  {\bibfnamefont {C.~R.}\ \bibnamefont {Harris}}, \bibinfo {author}
  {\bibfnamefont {A.~M.}\ \bibnamefont {Archibald}}, \bibinfo {author}
  {\bibfnamefont {A.~H.}\ \bibnamefont {Ribeiro}}, \bibinfo {author}
  {\bibfnamefont {F.}~\bibnamefont {Pedregosa}}, \bibinfo {author}
  {\bibfnamefont {P.}~\bibnamefont {{van Mulbregt}}},\ and\ \bibinfo {author}
  {\bibnamefont {{SciPy 1.0 Contributors}}},\ }\bibfield  {title} {\bibinfo
  {title} {{{SciPy} 1.0: Fundamental Algorithms for Scientific Computing in
  Python}},\ }\href {https://doi.org/10.1038/s41592-019-0686-2} {\bibfield
  {journal} {\bibinfo  {journal} {Nat. Methods}\ }\textbf {\bibinfo {volume}
  {17}},\ \bibinfo {pages} {261} (\bibinfo {year} {2020})}\BibitemShut
  {NoStop}%
\bibitem [{\citenamefont {Kamchatnov}\ and\ \citenamefont
  {Pitaevskii}(2008)}]{Kamchatnov2008}%
  \BibitemOpen
  \bibfield  {author} {\bibinfo {author} {\bibfnamefont {A.~M.}\ \bibnamefont
  {Kamchatnov}}\ and\ \bibinfo {author} {\bibfnamefont {L.~P.}\ \bibnamefont
  {Pitaevskii}},\ }\bibfield  {title} {\bibinfo {title} {{Stabilization of
  Solitons Generated by a Supersonic Flow of Bose-Einstein Condensate Past an
  Obstacle}},\ }\href {https://doi.org/10.1103/PhysRevLett.100.160402}
  {\bibfield  {journal} {\bibinfo  {journal} {Phys. Rev. Lett.}\ }\textbf
  {\bibinfo {volume} {100}},\ \bibinfo {pages} {160402} (\bibinfo {year}
  {2008})}\BibitemShut {NoStop}%
\bibitem [{Note1()}]{Note1}%
  \BibitemOpen
  \bibinfo {note} {See \protect \href
  {https://arxiv.org/src/2505.09553v4/anc/Supplemental_Material_I.mp4}
  {Supplemental Material~I} for a video presentation of the soliton
  motion.}\BibitemShut {Stop}%
\bibitem [{\citenamefont {Dormand}\ and\ \citenamefont
  {Prince}(1980)}]{Dormand1980}%
  \BibitemOpen
  \bibfield  {author} {\bibinfo {author} {\bibfnamefont {J.~R.}\ \bibnamefont
  {Dormand}}\ and\ \bibinfo {author} {\bibfnamefont {P.~J.}\ \bibnamefont
  {Prince}},\ }\bibfield  {title} {\bibinfo {title} {{A family of embedded
  Runge-Kutta formulae}},\ }\href
  {https://doi.org/10.1016/0771-050X(80)90013-3} {\bibfield  {journal}
  {\bibinfo  {journal} {J. Comput. Appl. Math.}\ }\textbf {\bibinfo {volume}
  {6}},\ \bibinfo {pages} {19} (\bibinfo {year} {1980})}\BibitemShut {NoStop}%
\bibitem [{Note2()}]{Note2}%
  \BibitemOpen
  \bibinfo {note} {See \protect \href
  {https://arxiv.org/src/2505.09553v4/anc/Supplemental_Material_II.mp4}
  {Supplemental Material~II} for a video presentation corresponding to
  Fig.~\ref {fig:2d:disk}.}\BibitemShut {Stop}%
\bibitem [{Note3()}]{Note3}%
  \BibitemOpen
  \bibinfo {note} {See \protect \href
  {https://arxiv.org/src/2505.09553v4/anc/Supplemental_Material_III.mp4}
  {Supplemental Material~III} for a video presentation corresponding to
  Fig.~\ref {fig:2d:drop}.}\BibitemShut {Stop}%
\bibitem [{Note4()}]{Note4}%
  \BibitemOpen
  \bibinfo {note} {See \protect \href
  {https://arxiv.org/src/2505.09553v4/anc/Supplemental_Material_IV.mp4}
  {Supplemental Material~IV} for a video presentation corresponding to
  Fig.~\ref {fig:2d:chain}.}\BibitemShut {Stop}%
\bibitem [{Note5()}]{Note5}%
  \BibitemOpen
  \bibinfo {note} {See \protect \href
  {https://arxiv.org/src/2505.09553v4/anc/Supplemental_Material_V.mp4}
  {Supplemental Material~V} for a video presentation corresponding to Fig.~\ref
  {fig:2d:evolution}.}\BibitemShut {Stop}%
\bibitem [{\citenamefont {Gavrilov}(2014)}]{Gavrilov2014.prb.b}%
  \BibitemOpen
  \bibfield  {author} {\bibinfo {author} {\bibfnamefont {S.~S.}\ \bibnamefont
  {Gavrilov}},\ }\bibfield  {title} {\bibinfo {title} {{Blowup dynamics of
  coherently driven polariton condensates}},\ }\href
  {https://doi.org/10.1103/PhysRevB.90.205303} {\bibfield  {journal} {\bibinfo
  {journal} {Phys. Rev. B}\ }\textbf {\bibinfo {volume} {90}},\ \bibinfo
  {pages} {205303} (\bibinfo {year} {2014})}\BibitemShut {NoStop}%
\bibitem [{\citenamefont {Gavrilov}(2022)}]{Gavrilov2022}%
  \BibitemOpen
  \bibfield  {author} {\bibinfo {author} {\bibfnamefont {S.~S.}\ \bibnamefont
  {Gavrilov}},\ }\bibfield  {title} {\bibinfo {title} {{Spin oscillations of a
  single-mode polariton system driven by a plane wave}},\ }\href
  {https://doi.org/10.1103/PhysRevB.106.045304} {\bibfield  {journal} {\bibinfo
   {journal} {Phys. Rev. B}\ }\textbf {\bibinfo {volume} {106}},\ \bibinfo
  {pages} {045304} (\bibinfo {year} {2022})}\BibitemShut {NoStop}%
\bibitem [{\citenamefont {Zibold}\ \emph {et~al.}(2010)\citenamefont {Zibold},
  \citenamefont {Nicklas}, \citenamefont {Gross},\ and\ \citenamefont
  {Oberthaler}}]{Zibold2010}%
  \BibitemOpen
  \bibfield  {author} {\bibinfo {author} {\bibfnamefont {T.}~\bibnamefont
  {Zibold}}, \bibinfo {author} {\bibfnamefont {E.}~\bibnamefont {Nicklas}},
  \bibinfo {author} {\bibfnamefont {C.}~\bibnamefont {Gross}},\ and\ \bibinfo
  {author} {\bibfnamefont {M.~K.}\ \bibnamefont {Oberthaler}},\ }\bibfield
  {title} {\bibinfo {title} {{Classical Bifurcation at the Transition from Rabi
  to Josephson Dynamics}},\ }\href
  {https://doi.org/10.1103/PhysRevLett.105.204101} {\bibfield  {journal}
  {\bibinfo  {journal} {Phys. Rev. Lett.}\ }\textbf {\bibinfo {volume} {105}},\
  \bibinfo {pages} {204101} (\bibinfo {year} {2010})}\BibitemShut {NoStop}%
\bibitem [{\citenamefont {Cominotti}\ \emph {et~al.}(2023)\citenamefont
  {Cominotti}, \citenamefont {Berti}, \citenamefont {Dulin}, \citenamefont
  {Rogora}, \citenamefont {Lamporesi}, \citenamefont {Carusotto}, \citenamefont
  {Recati}, \citenamefont {Zenesini},\ and\ \citenamefont
  {Ferrari}}]{Cominotti2023}%
  \BibitemOpen
  \bibfield  {author} {\bibinfo {author} {\bibfnamefont {R.}~\bibnamefont
  {Cominotti}}, \bibinfo {author} {\bibfnamefont {A.}~\bibnamefont {Berti}},
  \bibinfo {author} {\bibfnamefont {C.}~\bibnamefont {Dulin}}, \bibinfo
  {author} {\bibfnamefont {C.}~\bibnamefont {Rogora}}, \bibinfo {author}
  {\bibfnamefont {G.}~\bibnamefont {Lamporesi}}, \bibinfo {author}
  {\bibfnamefont {I.}~\bibnamefont {Carusotto}}, \bibinfo {author}
  {\bibfnamefont {A.}~\bibnamefont {Recati}}, \bibinfo {author} {\bibfnamefont
  {A.}~\bibnamefont {Zenesini}},\ and\ \bibinfo {author} {\bibfnamefont
  {G.}~\bibnamefont {Ferrari}},\ }\bibfield  {title} {\bibinfo {title}
  {{Ferromagnetism in an Extended Coherently Coupled Atomic Superfluid}},\
  }\href {https://doi.org/10.1103/PhysRevX.13.021037} {\bibfield  {journal}
  {\bibinfo  {journal} {Phys. Rev. X}\ }\textbf {\bibinfo {volume} {13}},\
  \bibinfo {pages} {021037} (\bibinfo {year} {2023})}\BibitemShut {NoStop}%
\bibitem [{\citenamefont {Keldysh}(2017)}]{Keldysh2017.en}%
  \BibitemOpen
  \bibfield  {author} {\bibinfo {author} {\bibfnamefont {L.~V.}\ \bibnamefont
  {Keldysh}},\ }\bibfield  {title} {\bibinfo {title} {{Coherent states of
  excitons}},\ }\href {https://doi.org/10.3367/UFNe.2017.10.038227} {\bibfield
  {journal} {\bibinfo  {journal} {Phys. Usp.}\ }\textbf {\bibinfo {volume}
  {60}},\ \bibinfo {pages} {1180} (\bibinfo {year} {2017})}\BibitemShut
  {NoStop}%
\end{thebibliography}
\end{document}